\documentclass[draft]{agujournal2019}
\usepackage{url}
\usepackage{lineno}
\usepackage[inline]{trackchanges}
\usepackage{soul}
\usepackage{amssymb}
\usepackage{amsmath}
\usepackage{siunitx}

\draftfalse

\journalname{Journal of Advances in Modeling Earth Systems (JAMES)}

\begin{document}

\title{Online learning in idealized ocean gyres}

\authors{James R. Maddison\affil{1}}

\affiliation{1}{School of Mathematics and Maxwell Institute for Mathematical
Sciences, The University of Edinburgh}

\correspondingauthor{James R. Maddison}{j.r.maddison@ed.ac.uk}

\begin{keypoints}
\item An end-to-end differentiable dynamical model and machine learning framework
\item Applied to learn missing eddy contributions in coarse resolution idealized ocean gyres
\item Online learning approach leads to coarse resolution calculations with restored mean flows and intrinsic variability
\end{keypoints}

\begin{abstract}
Ocean turbulence parameterization has principally been based on processed-based approaches, seeking to embed physical principles so that coarser resolution calculations can capture the net influence of smaller scale unresolved processes. More recently there has been an increasing focus on the application of data-driven approaches to this problem. Here we consider the application of online learning to data-driven eddy parameterization, constructing an end-to-end differentiable dynamical solver forced by a neural network, and training the neural network based on the dynamics of the combined hybrid system. This approach is applied to the classic barotropic Stommel-Munk gyre problem --  a highly idealized configuration which nevertheless includes multiple flow regimes, boundary dynamics, and a separating jet, and therefore presents a challenging test case for the online learning approach. It is found that a neural network which is suitably trained can lead to a coarse resolution neural network parameterized model which is stable, and has both a reasonable mean state and intrinsic variability. This suggests that online learning is a powerful tool for studying the problem of ocean turbulence parameterization.
\end{abstract}

\section*{Plain Language Summary}
The global ocean has large scale flows which are important for transporting quantities such as heat around the world. However when looked at more closely, on top of these large scale ocean flows there are also much smaller and very complicated flow features. Computer simulations of the ocean cannot capture all of these complicated small-scale features, and so instead we try to add their effects into the computer models. In this article we seek to use a machine learning approach to learn what the effect of these missing small scale flow features should be. We specifically apply `online learning', which monitors how well the model is doing and uses this to improve the model. We apply this to a very basic `ocean-in-a-box' model which captures some of the key principles of a wind driven ocean basin. The basic model is simple enough that we can study it in detail, but complicated enough that this is a difficult test for the online learning approach.

\section{Introduction}

The ocean is a highly turbulent fluid, and numerical ocean modeling requires the use of appropriate turbulence closures so as to represent the impact of unresolved small scales \cite{fox-kemper2019,hewitt2020}. Ocean parameterization efforts for large-scale and long-time numerical modeling have often focused on the ocean mesoscale, seeking to capture the net effect of this largest scale component of ocean turbulence, with the aim of more physically realistic larger-scale and longer-term calculations at lower cost and non-eddy-resolving resolution.

At coarse and non-eddy-resolving resolution the ocean mesoscale is often parameterized using a form of the Gent-McWilliams parameterization \cite{gent1990,gent1995}. Extensions of the Gent-McWilliams parameterization have been developed, and in particular there has been a focus on maintaining energetic consistency by imposing energetic constraints on the parameterization \cite{cessi2008,eden2008,jansen2015_a,mak2018}.

As computational power has increased there has been an increasing move towards ocean parameterization in the partially eddy resolving `gray zone' \cite{christensen2022}, where the largest scales of the turbulent field can be captured explicitly. A key focus has been on backscatter capable parameterizations \cite{jansen2014,grooms2015,jansen2015_b,bachman2019,juricke2020,yankovsky2024}, which can tackle an excessive numerical drain of energy at smaller scales, and which can be motivated by the physical principle of energy backscatter by barotropic stability. As numerical methods improve and computational power increases further (see e.g. the recent \citeA{silvestri2024-preprint}) it is reasonable to expect that parameterizations for increasingly small-scale processes will continue to need to be developed.

Alongside these process-based approaches, in recent years there has been an increasing effort to apply data-driven methods. This has principally focused on offline diagnostic approaches: finding some means of diagnosing missing `eddy' terms, and then applying data analysis or machine learning techniques to find ways to predict those missing eddy terms given knowledge only of low resolution fields. An oceanographic data-driven approach of this type is described in \citeA{portamana2014}, for an idealized quasigeostrophic model of ocean gyres. Here, coarse graining of model fields and diagnosed equation terms was used to test relationships between a missing eddy term and proposed parameterizations. However the proposed parameterization which resulted involves an indefinite Helmholtz operator, causing difficulties when applied in a deterministic prognostic calculation \cite{zanna2017}.

This type of data-driven methodology naturally extends to an `offline' learning approach, with reference data obtained from a high resolution model, and the data used to train a machine learning system \cite{bolton2019,guillaumin2021,ross2023,srinivasan2024} or for equation discovery \cite{zanna2020}.

The central issue encountered in these offline approaches is that having a small but finite error in a tendency does not mean that there is a small error in a solution to dynamical equations. Small contributions might be missed in offline training, but nevertheless be essential for stability. This might particularly be expected to be the case in systems which backscatter at one scale and dissipate at another, as there is no clear reason for the dissipation to correspond to tendencies of large magnitude.

More recently `online' learning approaches have been applied. Also known as `hybrid' or `a posteriori' approaches, these methods consider a numerical solver coupled to a neural network. Examples appear in finite element modeling using FEniCS \cite{berg2018-preprint,mitusch2021}, ADCME \cite{huang2020}, and Firedrake \cite{bouziani2023}. Online learning has recently been applied with notable success for atmospheric general circulation modeling in \citeA{kochkov2024}, has been applied for fluid flow modeling \cite{sirignano2020,um2020,kochkov2021,list2022}, and for fluid super-resolution \cite{page2024-preprint}. An application of online learning to an ocean relevant single layer quasi-geostrophic model is described in \citeA{frezat2022}, and an application to a two-layer quasi-geostrophic baroclinic instability problem is described in \citeA{yan2024-preprint}.

The core change in an online learning approach is to define some input to a model in terms of a neural network, and then to define a loss in terms of the output of the combined neural network and dynamical solver. For example in the quasi-geostrophic application in \citeA{frezat2022} the model is forced with an additional neural network defined term, and the solution of the dynamical equations at lower resolution is compared with a higher resolution reference. Such an online approach directly tackles the stability issues previously encountered in offline approaches -- if the parameterized model is unstable, then this can be expected to lead to an increased loss. In \citeA{frezat2022} an online learning approach is indeed found to lead to improved stability.

An online learning approach presents significant additional technical challenge as compared against an offline approach. Evaluation of a loss, comparing evolved solutions with reference data, requires the solution of dynamical equations, increasing the computational cost. Worse, evaluation of the gradient of the loss, required for efficient training, requires the ability to differentiate through the dynamical model and solve the resulting discrete adjoint equations. The required derivative information can be obtained using automatic differentiation \cite{griewank2008,baydin2018} (also referred to as `algorithmic differentiation' or `autodiff'), but this still requires the implementation of the entire dynamical model in a form which is compatible with an automatic differentiation tool. Automatic differentiation has been applied with great success to ocean modeling in the differentiable MITgcm code \cite{heimbach2002,heimbach2005}, and is a key technology in the well-establish use of variational data assimilation (e.g. as applied for ocean state estimation with ECCO, \citeA{forget2015}). The key change is that, whereas variational data assimilation is used to infer fields in discrete spaces, online learning is used to infer an operator mapping \emph{between} discrete spaces.

This article describes the application of an online learning approach in an idealized, but nevertheless challenging, ocean relevant configuration. Specifically a classic wind-forced barotropic Stommel-Munk model in a double gyre configuration is considered. This configuration is sufficiently simple that a high resolution eddy resolving reference calculation can be performed without significant computational cost, but nevertheless includes a number of challenging elements: the domain is bounded, the flow is anisotropic and inhomogeneous, the flow exhibits multiple regimes in different parts of the domain, and the flow includes the dynamics of a separating jet.

At coarse resolution the model is driven by an extra neural network term, allowing neural networks to be trained by comparing dynamically evolved solutions with reference data. The dynamical equations are implemented using Python with JAX \cite{frostig2018}, and the neural networks are implemented with Keras \cite{keras2024} using the JAX backend, providing an end-to-end differentiable system suitable for online learning, and with the side-benefit of allowing the entire system to run efficiently on a GPU. This efficiency significantly aids in both the generation of reference data and neural network training

This article represents a first attempt to apply online learning in a bounded and inhomogeneous ocean configuration, and seeks to determine whether online learning can be used for an improved representation of the mean flow and variability. A secondary aim is to assess the ability of online learning to learn a form for explicit numerical dissipation at coarse resolution. While this article is not primarily focused on issues  generalizability, some `out-of-sample' tests are performed.

A practical parameterization must be both generalizable to new problems of interest, and cheap enough to use -- versus, say, simply increasing resolution. This article makes use of deep convolutional neural networks consisting of multiple filters in each layer, and applies these with a very simple, small, and efficient dynamical model. The specific neural network architecture tested here may not be competitive in terms of practical performance. The purpose, rather, is to test the viability of online learning when applied to ocean turbulence problems.

While performance and stability is variable, the key findings are that a coarse resolution model driven by a neural network, after appropriate online training, is able to run stably, is able to maintain a reasonable mean flow state, and is moreover able to maintain a reasonable level of intrinsic variability. Remarkably, isolated eddies are observed evolving in coarse resolution simulations, despite the very low resolution used.

The article proceeds as follows. In section \ref{sect:methods} the end-to-end differentiable Python JAX model is introduced, and the methodology used for online learning is described. Section \ref{sect:results} investigates the performance of trained neural networks for this problem, focusing on the ability of the resulting combined dynamical model and neural network system to capture the mean, variability, and to remain numerically stable. Generalizability is briefly considered in section \ref{sect:generalizability}. The article concludes in section \ref{sect:conclusions}.

\section{Methodology}\label{sect:methods}

\subsection{Equations and discretization}

A classic Stommel-Munk double gyre configuration on a $\beta$-plane is considered (see chapter 19 of \citeA{vallis2017}). With a rigid lid approximation and using a stream-function--vorticity formulation the dynamical equations are
\begin{equation*}
  \partial_t \zeta + u \cdot \nabla ( \zeta + \beta y )
    = -r \zeta + \nu \nabla^2 \zeta + \frac{1}{\rho_0 D} \nabla^\perp \cdot \tau,
\end{equation*}
\begin{equation*}
  \zeta = \nabla^2 \psi.
\end{equation*}
Here $\zeta$ is the relative vorticity, $\psi$ is the stream function, $u = \nabla^\perp \psi$ is the velocity, $\beta$ is the meridional derivative of the planetary vorticity, $r$ is a linear drag coefficient, $\nu$ is a Laplacian viscosity coefficient, $\rho_0$ is the density, $D$ is the depth, and $\tau$ is the surface wind stress. $x$ and $y$ are the zonal and meridional coordinates, and $t$ is time. The equations are solved in a domain $x \in [ -L, L ]$ and $y \in [ -L, L ]$. Physical parameters and the wind stress are based on \citeA{marshall2010}, although here a Laplacian viscosity is used and a linear drag is included. Parameters are listed in Table~\ref{tab:physical_parameters}, with a wind stress
\begin{equation}\label{eqn:wind_stress}
  \tau = \tau_0 \cos \left( \frac{ \pi y}{L} \right) \left( \begin{array}{c} 1 \\ 0 \end{array} \right),
\end{equation}
where $\tau_0$ is the maximum surface wind stress magnitude. No-normal-flow and free-slip boundary conditions are applied, with $\psi = 0$ and $\zeta = 0$ on all boundaries. The use of free-slip boundary conditions simplifies the use of implicit timestepping for the dissipation terms, and also later simplifies the application of convolutional neural networks, where zero padding is applied.

\begin{table}
\caption{Physical parameters. $L$, $\beta$, $\tau_0$, and $D$ are as in \citeA{marshall2010}.}\label{tab:physical_parameters}
\begin{center}\begin{tabular}{c|c|c}
  Parameter & Symbol & Value \\
  \hline
  Domain size & $2 L$ & $4000$~\unit{\km} \\
  Meridional derivative of planetary vorticity & $\beta$ & $2 \times 10^{-11}$~\unit{\per\m\per\s} \\
  Laplacian viscosity coefficient & $\nu$ & $10$~\unit{\m\squared\per\s} \\
  Linear drag coefficient & $r$ & $10^{-7}$~\unit{\per\s} \\
  Maximum wind stress magnitude & $\tau_0$ & $0.1$~\unit{\N\per\m\squared} \\
  Density & $\rho_0$ & $10^3$~\unit{\kg\per\m\cubed} \\
  Depth & $D$ & $500$~\unit{\m}
\end{tabular}\end{center}
\end{table}

The equations are discretized in space using the Arakawa Jacobian \cite{arakawa1966} for the advection term, second order centred differencing for other spatial derivatives, and interpolation of the wind stress curl. This leads to a semi-discrete form
\begin{equation*}
  d_t \tilde{\zeta} = -N ( \tilde{\psi}, \tilde{\zeta} + \beta \tilde{y} ) + L ( \tilde{\zeta} ) + \tilde{Q}.
\end{equation*}
Here $\tilde{( \ldots )}$ denotes a vector of degrees-of-freedom for a discretized field. $N ( \cdot, \cdot )$ denotes the discretization for the advection term, $L ( \cdot )$ is a linear operator denoting the discretization of the linear drag and Laplacian viscosity terms, and $\tilde{Q}$ is the discrete wind stress forcing term. A neural network forced system is constructed by adding an additional term
\begin{equation*}
  d_t \tilde{\zeta} = -N ( \tilde{\psi}, \tilde{\zeta} + \beta \tilde{y} ) + L ( \tilde{\zeta} ) + \tilde{Q} + \alpha_\text{output} \mathcal{F}_\theta ( \alpha_\text{input} \tilde{\zeta} ),
\end{equation*}
where $\mathcal{F}_\theta ( \cdot )$ represents a neural network with weights $\theta$. Normalization factors
\begin{align*}
  \alpha_\text{input} & = \frac{1}{\left| \beta \right| L}, \\
  \alpha_\text{output} & = \frac{| \tau_0 | \pi}{\rho_0 D L},
\end{align*}
are used to non-dimensionalize the neural network input and output.

A fully discrete system is reached by applying a CNAB2 time discretization \cite{ascher1995}, with the linear drag and Laplacian viscosity terms represented by $L ( \cdot )$ treated implicitly using a Crank-Nicolson discretization, and the remaining terms (including the neural network term) treated explicitly using a second order Adams-Bashforth discretization. On the first timestep a CNAB1 discretization is used, with a forward Euler discretization applied for the explicit terms. Elliptic problems are solved using Fast Fourier Transforms. Online training is performed in single precision, and all other calculations are performed in double precision.

\subsection{Implementation}

The dynamical model is implemented in Python using the JAX library. Neural networks are implemented using the Keras library using the JAX backend. Crucially, since both the dynamical model and the neural networks are implemented using JAX, the combined system is end-to-end differentiable and runs on a GPU.

\begin{figure}
 \begin{center}\includegraphics[height=0.55\textwidth]{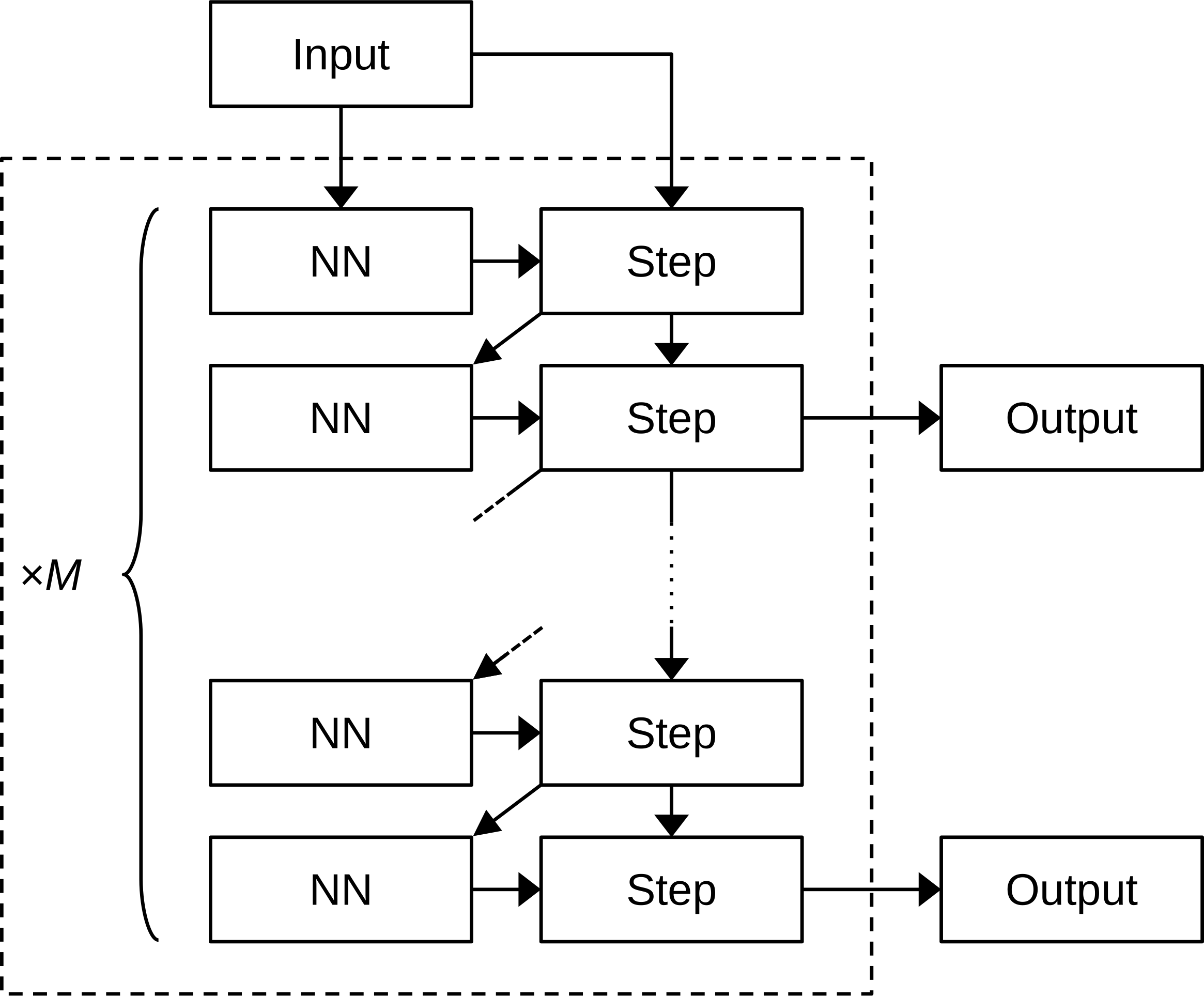}\end{center}
  \caption{A \texttt{Dynamics} layer. The layer takes as input an initial relative vorticity, and steps the model while evaluating the neural network (NN), periodically adding the current relative vorticity to the output. Here the \texttt{Dynamics} layer takes $2 M$ steps and outputs every $2$ steps -- this is the configuration later used in this article, where the \texttt{Dynamics} layer outputs every hour, with timesteps of $30$~minutes, and integrates over a window of $M = 24 w$~hours. In practice additional pre-processing is applied to the input, here so that the input relative vorticity satisfies the boundary conditions.}\label{fig:dynamics_layer}
\end{figure}

In order to train a neural network, the entire combined dynamical model and neural network system is itself defined to be a custom Keras \texttt{Dynamics} layer. The custom layer takes, as input, an initial condition for the discrete relative vorticity field $\tilde{\zeta}$, and timesteps the numerical model (evaluating the neural network on each timestep), while periodically appending the current relative vorticity field to the layer outputs. This is illustrated schematically in Figure~\ref{fig:dynamics_layer}. A \texttt{Dynamics} layer can conceptually itself be considered to be a recurrent network, with a nested neural network defining the forcing term and a fixed layer defining the timestep. The \texttt{Dynamics} layer is used to define an outer Keras model mapping inputs (initial conditions) to outputs (the dynamical trajectory evaluated from those initial conditions) which can then be trained. The current implementation allows any neural network to be embedded within the dynamical model, so long as evaluation of the network does not change its state or have other side effects other than to update the dynamical model (e.g. batch normalization can not currently be used).

Note that the dynamical model therefore supports two types of coupling with neural networks: a neural network can be embedded within and used to force the dynamical model, and the dynamical model can be embedded within a more complicated network. Only the former is applied in this article.

\subsection{Data generation}

A dataset is generated by running the dynamical model, with no neural network forcing, at a high resolution. Specifically the model is run with a uniform resolution $2049 \times 2049$ grid, corresponding to a grid spacing of $1.95$~\unit{\km}, with a Laplacian viscosity of $\nu = 10$~\unit{\m\squared\per\s}. The Munk length for this configuration is $L_m = ( \nu / \beta )^{1/3} = 7.94$~\unit{k\m} and the Stommel length is $L_s = r / \beta = 5$~\unit{\km}. A timestep of $\Delta t = 2$~minutes is used.

A pseudorandom perturbation is applied to the initial relative vorticity field, and then the model is integrated for a spinup period of $12$~years (where in this article all units of years refer to common years of $365$~days). After the spinup period the model is integrated for a further $12$~years for data generation. Within the data generation window, and every hour, a filtered and coarse grained relative vorticity field is generated by first applying a Gaussian filter with a width (specifically a standard deviation) of $62.5$~\unit{\km}, and then downsampling onto a uniform resolution $65 \times 65$ grid with grid spacing $62.5$~\unit{\km}, leading to a filtered and coarse grained relative vorticity field on a grid with a grid spacing of $62.5$~\unit{\km}. This leads to a total of $N_\text{data} = 105121$ hourly values for a filtered and coarse grained relative vorticity.

The Gaussian filter as used here, with width equal to the target grid scale, is as in \citeA{zanna2020}. The choice of filtering and coarse graining operator is found to be an important element when applying offline learning in \citeA{ross2023}. See also \citeA{frezat2022} where the use of a Gaussian filter or a sharp spectral cutoff is compared in online learning.

An important detail in online learning, as it is applied here, is that the training data is used to both define a loss, and also to \emph{initialize} coarse resolution calculations during training. Given the factor $32$ increase in grid spacing applied here, together with the fine scales which appear in the vorticity field in the high resolution reference calculation, we choose a filtering and coarse graining operator which reduces grid scale noise on the coarse resolution grid.

\subsection{Training and validation data sets}

A \texttt{Dynamics} layer maps a single input value for the relative vorticity to a number of output values defined by integrating the dynamical model with the neural network forcing. The training set should therefore similarly consist of a set of input initial conditions, together with a set consisting of later output states associated with each initial condition.

The training set is defined using the first $80$\% of the $12$~year record of filtered and coarse grained relative vorticity. A set of windows of a given length -- each of length, say $w$~days -- is defined. Within each window the first value for the filtered and coarse grained relative vorticity defines an input, and the later values define an associated output.

Specifically for a batch size of $b$ the first
\begin{equation*}
    \left\lfloor \frac{\frac{4 N_\text{data}}{5} - M - 1}{M b} \right\rfloor M b + M + 1
\end{equation*}
hourly values in the $12$~year dataset are assigned to the training set, where $N_\text{data}$ is the total number of hourly outputs and $M = 24 w$ is the number of hourly outputs within each window. At the start of training, and at the end of each epoch, a pseudorandom initial index $i_0 \in [ 0, M ]$ is chosen. Each window in the training set then ranges over hour indices $i \in [ i_0 + n M, i_0 + (n + 1) M ]$ for non-negative integer $n$, with the first index defining an input, and the remaining indices defining the associated output. The use of a pseudorandom initial index $i_0$ provides a richer dataset, ensuring that different possible evolution periods are considered during training. The order of the windows is also shuffled at the start of training and at the end of each epoch.

The final $20$\% of the data is assigned to a validation. Specifically windows in the validation set have hour indices $i \in [i_1 + n M, i_1 + (n + 1 ) M ]$ for non-negative integer $n$, where $i_1 = \left\lfloor 4 N_\text{data} / 5 \right\rfloor$.

\subsection{Neural network architecture}

A residual neural network (ResNet) is considered. A base network based on the architecture used in \citeA{frezat2022} is initially constructed, although is smaller, with $7$ convolutional layers each with $32$ filters of width $5$ and with ELU activation functions \cite{clevert2016-preprint}. Skip connections are then added from the input layer, and from all subsequent layers. Each skip connection permits a trainable scaling factor and trainable scalar bias, and the output is then formed by adding the results -- so that the output is a linear combination of the input and the output of each layer, plus some scalar bias. The rationale here is that earlier layers might include more local information -- for example representing lower order differential operators -- and that it might be beneficial to include this information more directly in the output. The ResNet has a total of $154857$ trainable weights.

\begin{figure}
  \begin{center}\begin{tabular}{c}
    \includegraphics[height=0.55\textwidth]{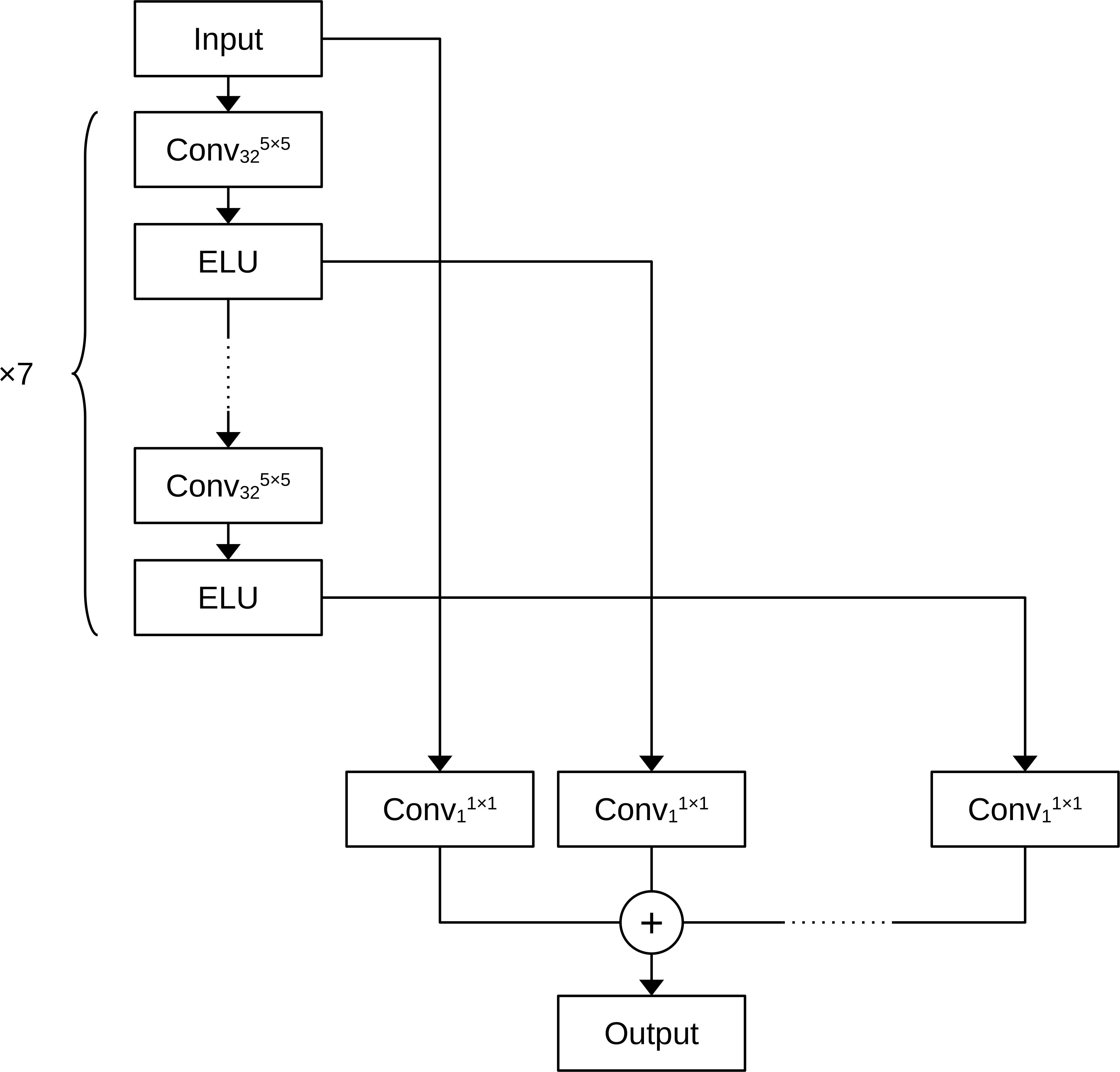} \\
    \qquad \quad ResNet
  \end{tabular}\end{center}
  \caption{ResNet neural network architecture considered in this article. Convolution layer notation is as in Figure~2 of \citeA{frezat2022}. In the ResNet, single filter and width $1$ convolution layers are used to construct a linear combination of the input and convolutional layer outputs, with a bias. All convolutional layers include a bias, and all convolution layers with width greater than $1$ use padding.}\label{fig:cnn}
\end{figure}

The considered neural network architecture is illustrated schematically in Figure~\ref{fig:cnn}.

\subsection{Coarse resolution configuration and training}

The \texttt{Dynamics} layer is configured at a much lower resolution, matching the $65 \times 65$ grid of the filtered and coarse grained data. A larger timestep of $\Delta t = 30$~minutes is used, but all other parameters are unchanged. Crucially the explicit Laplacian viscosity of only $\nu = 10$~\unit{\m\squared\per\s} is retained at coarse resolution. Such a low explicit viscosity is unsuitable for long simulations at this low resolution -- the model, with no neural network contribution, is in fact stable in this configuration, but has significant grid scale noise. Retaining a very low explicit viscosity in the coarse resolution dynamical model leaves the neural network to infer an appropriate form for the dissipation. That is, we are starting with an inherently under-dissipated model, and leave the neural network to apply an appropriate form of dissipation.

A loss is defined
\begin{equation*}
  J_\text{loss} = 10^4 \frac{1}{\beta^2 L^2} \frac{1}{4 L^2} \frac{1}{M} \frac{1}{N_\text{windows}} \sum_{i \in \left\{ 1, \ldots, M \right\}} \sum_{j \in \left\{ 1, \ldots, N_\text{windows} \right\}} ( \tilde{\zeta}_{i,j} - \tilde{\zeta}_{C,i,j} )^T W \left( \tilde{\zeta}_{i,j} - \tilde{\zeta}_{C,i,j} \right).
\end{equation*}
Here $\tilde{\zeta}_{i,j}$ is the degree-of-freedom vector for the value for hour $i$ in window $j$, with a total of $N_\text{windows}$ considered output windows each with $M = 24 w$ hourly outputs. $\tilde{\zeta}_{C,i,j}$ is the degree-of-freedom vector for the corresponding filtered and coarse grained relative vorticity. $W$ is here set equal to a diagonal matrix with elements equal to $\Delta x_C^2$ for interior nodes and zero for other nodes, where $\Delta x_C$ is the coarse resolution grid spacing. With this definition the loss is a scaled squared $L^2$ mismatch, excluding boundary nodes. The factor $10^4$ is included with the intention of improving training, but the value of this factor, and optimizer parameters, were not investigated in detail.

Training is performed using the AdamW optimizer \cite{loshchilov2019-preprint}, with Keras default training parameters. A batch size of $10$ is used. Training is performed over increasing window lengths, starting from a length of $w = 1$~day and increasing to $w = 60$~days in increments of $1$~day. Optimization for $5$ epochs is performed at each window width, with the neural network state retained from the previous window, and with the optimizer re-initialized when changing window length. This is analogous to the training approach in \citeA{frezat2022} and \citeA{kochkov2024}, but note the re-initialization of the optimizer here.

\section{Results}\label{sect:results}

\subsection{Control}

A control calculation is first considered, using the coarse resolution $65 \times 65$ grid with no neural network forcing. The model is integrated for a spinup of $12$~years, and then mean quantities are computed by time averaging over a further $12$~years. Throughout this article $\overline{\left( \ldots \right)}$ denotes a $12$~year time average after a $12$~year spinup, and $\left( \ldots \right)'$ denotes a deviation from this time average. Time averages for the $2049 \times 2049$ reference calculation are computed using values of fields evaluated every hour, and time averages for other calculations are computed using fields evaluated every timestep, with the time averages computed using composite trapezoidal rule integration.

The instantaneous potential vorticity field $q = \zeta + \beta y$, at the end of the full $24$~year simulation of the control calculation, is shown in Figure~\ref{fig:resnet_pv_4}. While the coarse resolution calculation is able to run stably with the unsuitably low Laplacian viscosity of $10$~\unit{\m\squared\per\s}, the results are clearly polluted with significant grid scale noise. The mean transport stream function for the coarse resolution $65 \times 65$ control calculation is shown in Figure~\ref{fig:resnet_mean_psi}. Compared with the high resolution $2049 \times 2049$ reference the mean jet is poorly represented at coarse resolution, with a significantly reduced eastward extent.

In this barotropic jet configuration the flow exhibits properties of barotropic instability in the upstream jet region, and then the flow transitions in the downstream region, exhibiting properties of barotropic stability and with eddy energy backscattering to drive the mean flow -- see \citeA{waterman2013} for a discussion. The reduced eastward extent of the jet in the coarse resolution calculation suggests that these mechanisms -- and in particular the downstream forcing of the jet by eddy backscatter -- are incorrectly represented at coarse resolution.

\subsection{Residual neural network (ResNet)}

\begin{figure}
  \begin{center}\begin{tabular}{c}
  \includegraphics[height=0.5\textwidth]{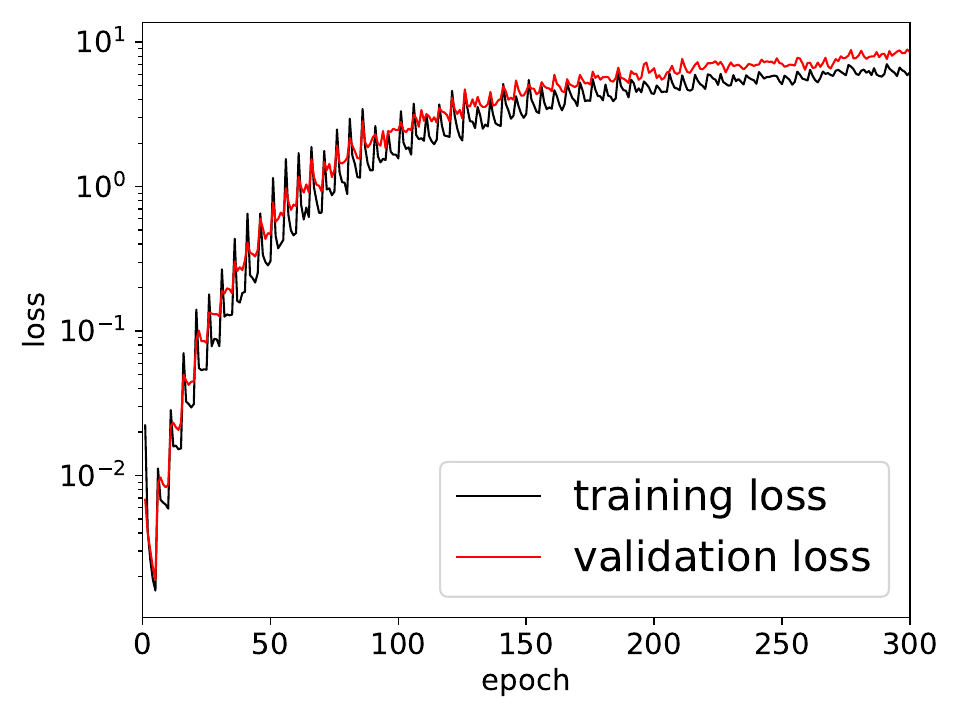} \\
  \qquad ResNet
  \end{tabular}\end{center}
  \caption{Training and validation loss during training of the ResNet, as reported by Keras. Every $5$ epochs the window length is increased and the optimizer re-initialized. Only the loss after each epoch is shown.}\label{fig:training}
\end{figure}

The residual neural network (ResNet) was also trained using the filtered and coarse grained high resolution dataset, leading to a further $60$ trained neural networks associated with window lengths of $1$--$60$ days. The training and validation loss as reported by Keras during training are shown in Figure~\ref{fig:training}. The loss steadily increases as the window length is increased.

The ResNet is used prognostically to form a ResNet parameterized model with a $65 \times 65$ grid. The two shortest window lengths at $w = 1$~day and $w =2$~days led to numerical instability, but all other cases led to stable calculations.

\begin{figure}
  \begin{center}\begin{tabular}{ccc}
    \includegraphics[height=0.3\textwidth]{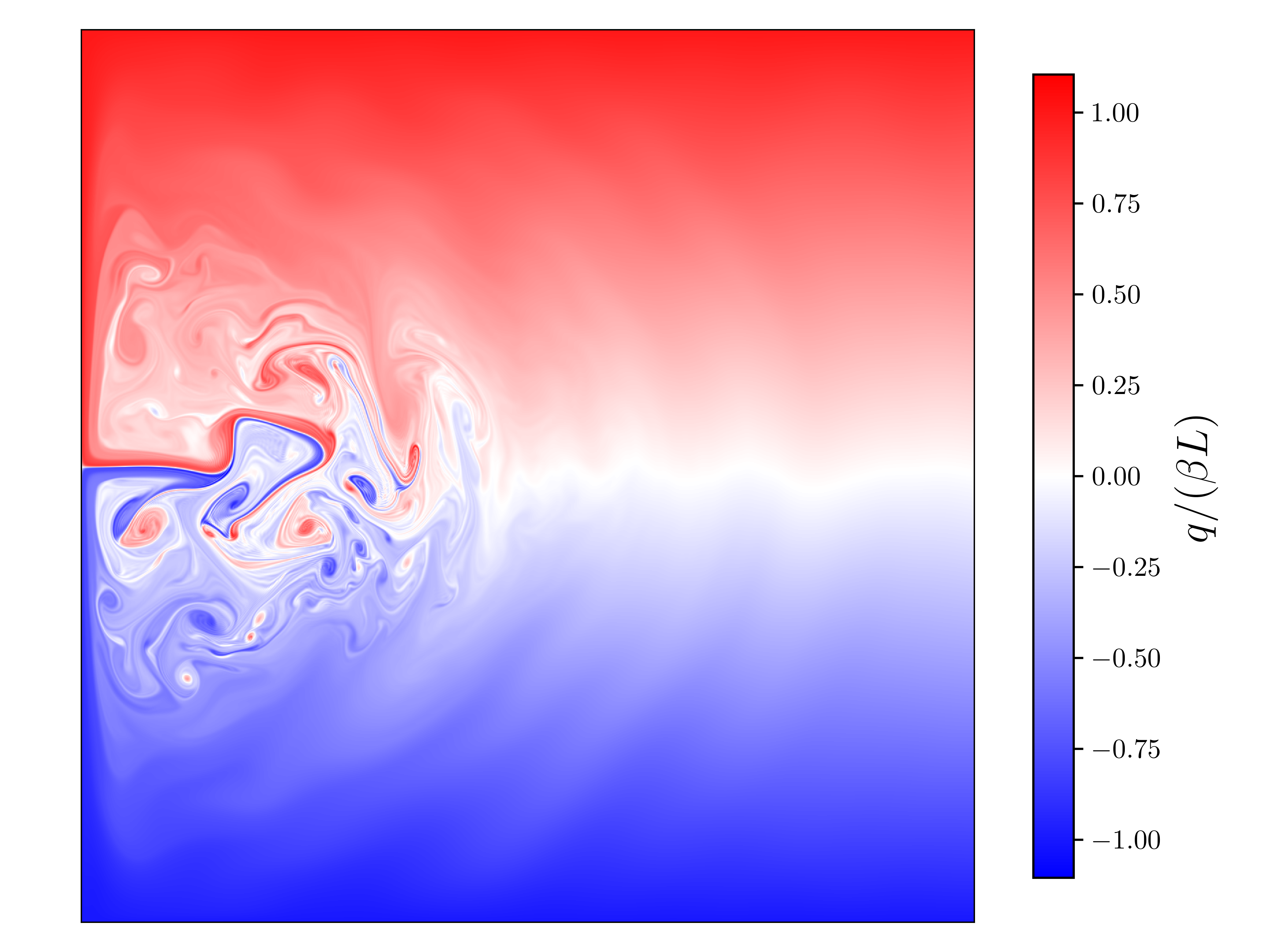}
    & &
    \includegraphics[height=0.3\textwidth]{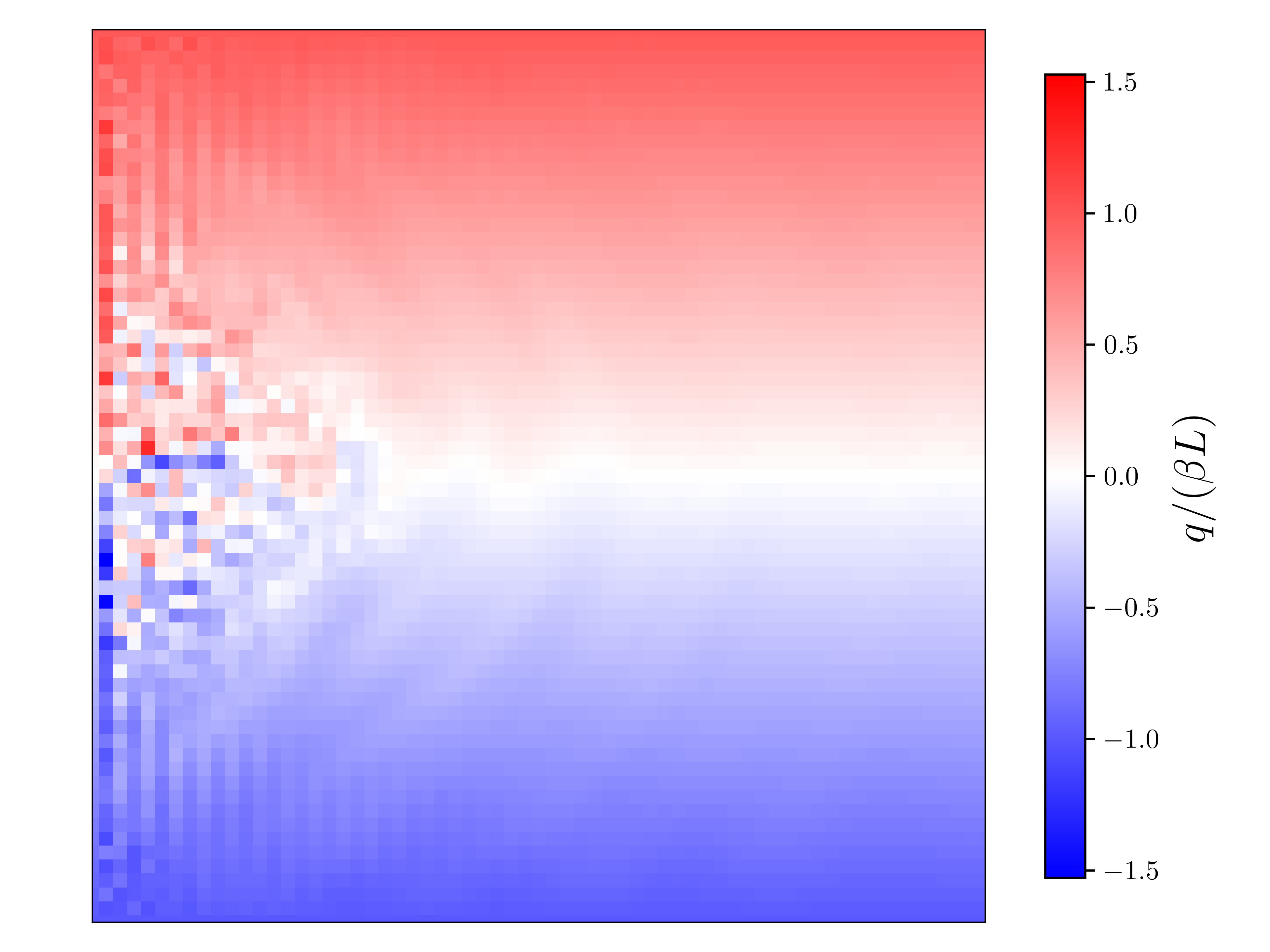} \\
    $2049 \times 2049$ \qquad\qquad
    & &
    $65 \times 65$ \qquad\qquad
  \end{tabular}
  \begin{tabular}{c}
    \includegraphics[height=0.3\textwidth]{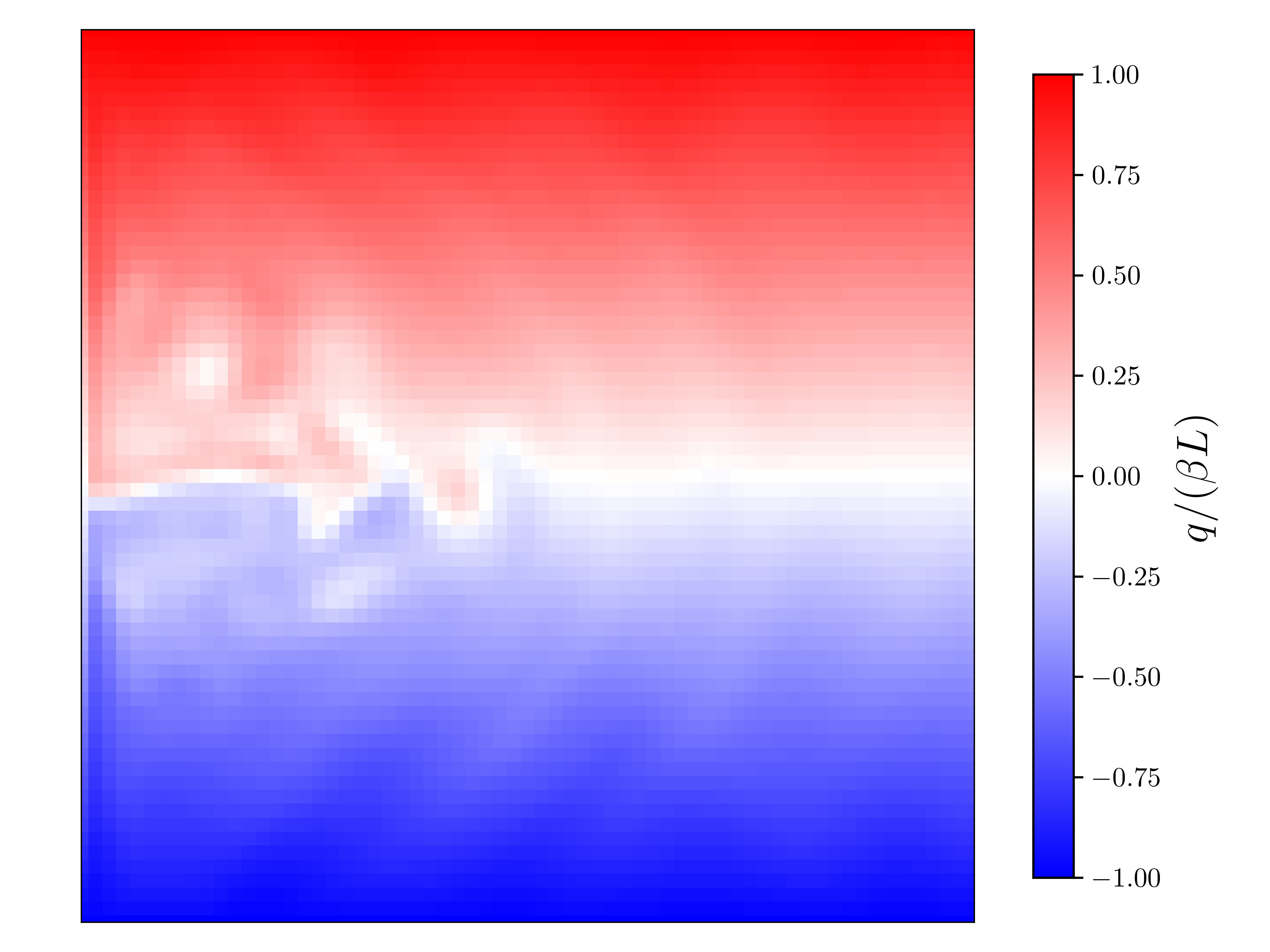} \\
    $65 \times 65$ + ResNet, $w = 4$~days \qquad\qquad
    \\
  \end{tabular}\end{center}
  \caption{Instantaneous potential vorticity $q + \beta y$, normalized by $\beta L$, after $24$~years, for the high resolution $2049 \times 2049$ reference, the coarse resolution $65 \times 65$ control, and the ResNet parameterized model after training up to a window length of $w = 4$~days. Note that different color scales are used.}\label{fig:resnet_pv_4}
\end{figure}

Figure~\ref{fig:resnet_pv_4} shows the instantaneous potential vorticity at the end of the $24$~year simulation after training up to a window length of $w = 4$~days. Remarkably coherent eddies can be observed despite the coarse resolution. A notable property is that the potential vorticity field appears smooth, despite the use of a very low explicit viscosity. That is, it appears that the neural network has learned to apply a dissipation to remove smaller scale noise, while also allowing the development of eddying dynamics on resolvable scales.

\begin{figure}
  \begin{center}\begin{tabular}{ccc}
    \includegraphics[height=0.3\textwidth]{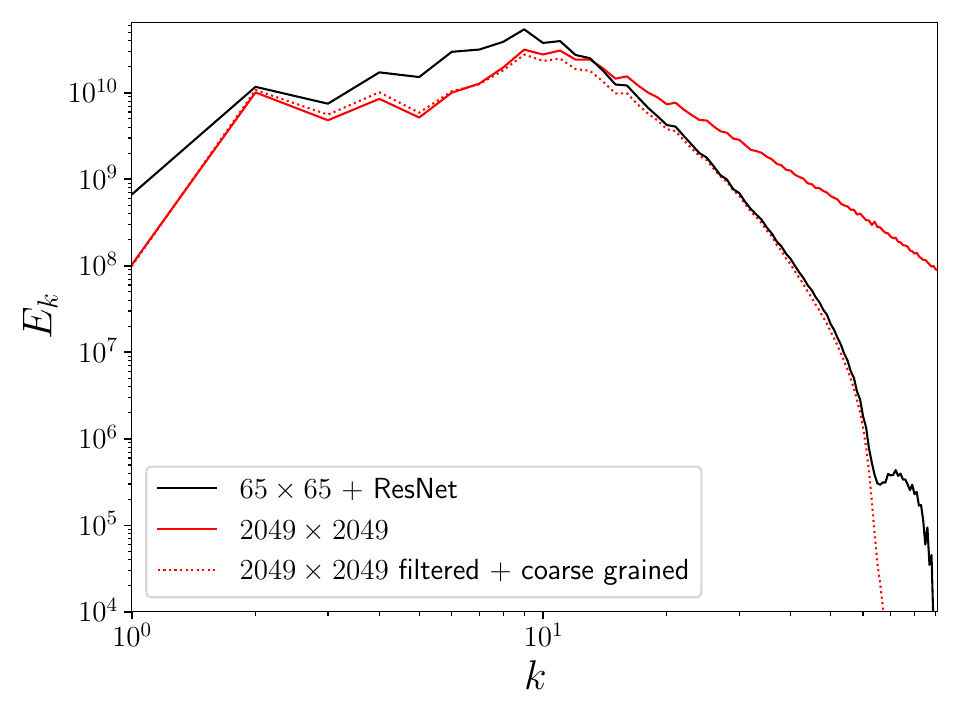}
    & &
    \includegraphics[height=0.3\textwidth]{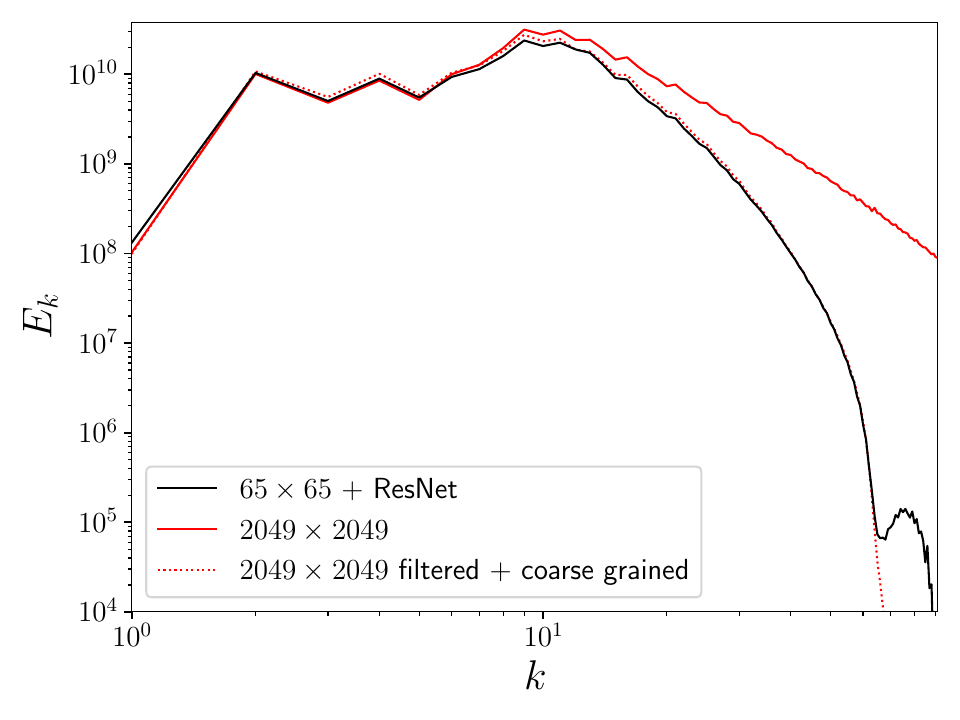} \\
    $w = 3$~days
    & &
    $w = 4$~days \\
    & & \\
    \includegraphics[height=0.3\textwidth]{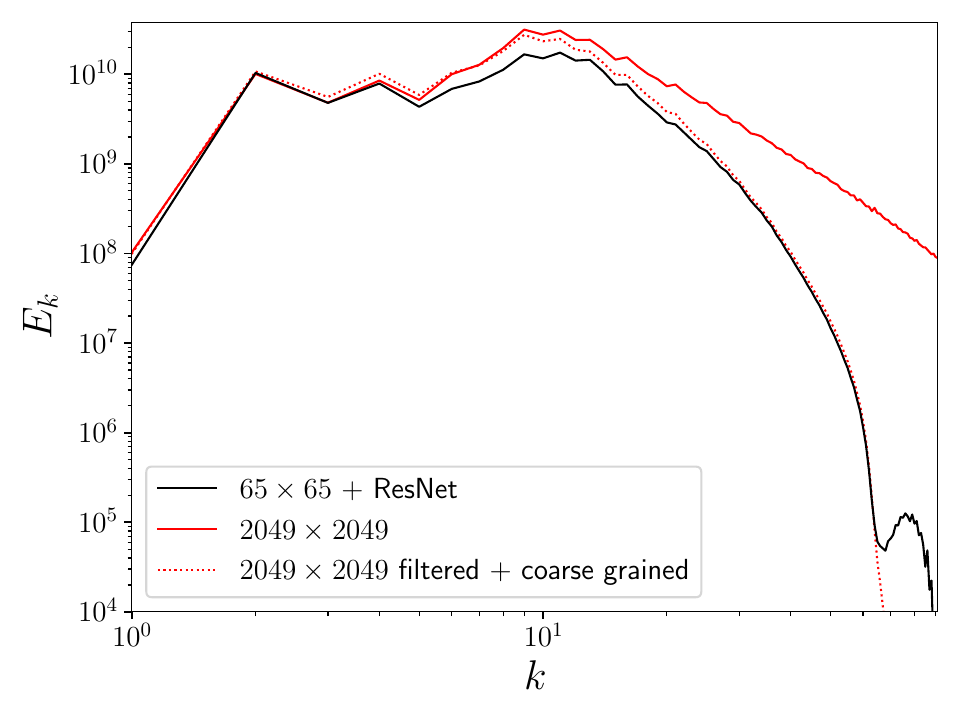}
    & &
    \includegraphics[height=0.3\textwidth]{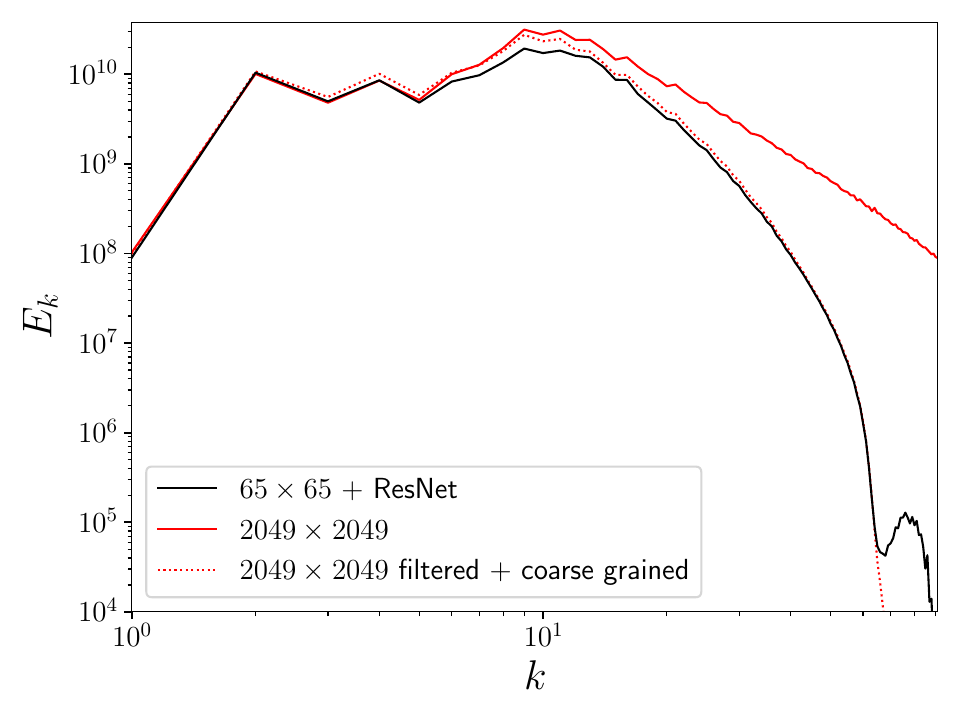} \\
    $w = 5$~days
    & &
    $w = 6$~days
  \end{tabular}\end{center}
  \caption{Time-averaged kinetic energy spectra for the ResNet parameterized model after training to different window lengths $w = 3$ to $w = 6$~days (black) with the time-averaged kinetic energy spectrum for the high resolution $2049 \times 2049$ reference (red) and filtered and coarse grained data (dotted red). $E_k$ has units \unit{\m^4\per\s\squared}, is computed from the discrete sine transform of the stream function, and is defined so that up to discretization error the sum of the full spectrum is the domain integral of the time-averaged kinetic energy. The wavenumber $k$ is normalized, with $k = 1$ corresponding to a wavelength of $4 L$.}\label{fig:resnet_ke_spectrum}
\end{figure}

Kinetic energy spectra, computed from the discrete sine transform of the stream function, are shown in Figure~\ref{fig:resnet_ke_spectrum}. Note that here and throughout this article `energy' refers to the `energy per unit mass'. After training up to a window length of $w = 3$~days the ResNet parameterized model exhibits excessive large scale energy. However after training up to a window length of $w = 4$~days the ResNet parameterized model spectrum rather closely matches that of the filtered and coarse grained data. There is some modest build up of energy near the grid scale, but the solution remains dramatically smoother than for the $65 \times 65$ control. Much more small scale energy is seen at some longer window lengths, which is particularly apparent at window length $w = 58$~days (not shown).

\begin{figure}
  \begin{center}\includegraphics[height=0.5\textwidth]{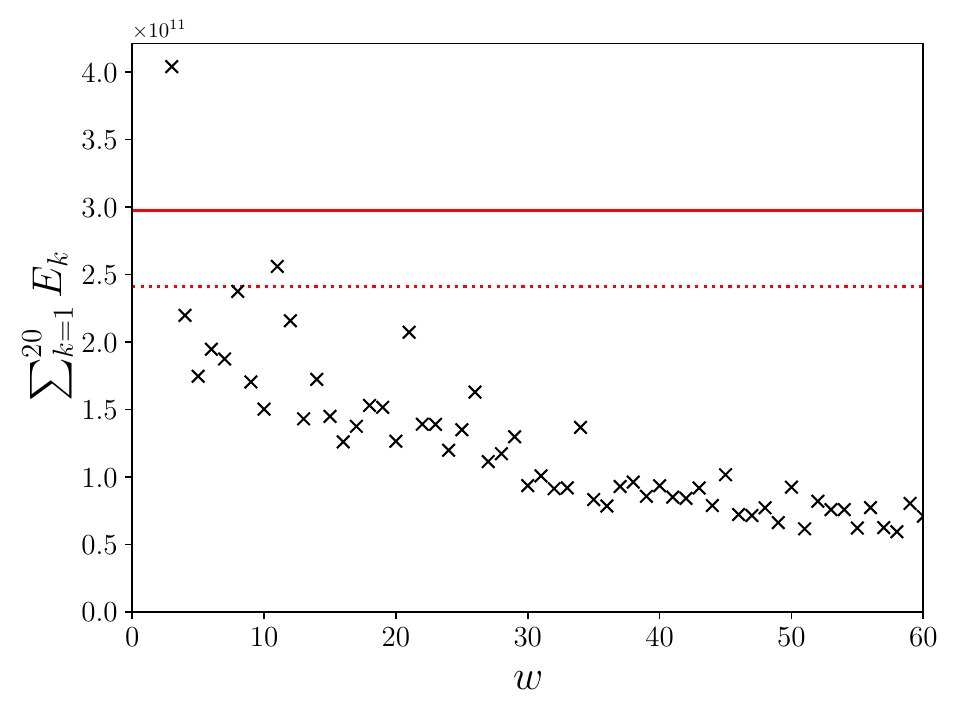}\end{center}
  \caption{Large scale kinetic energy in the ResNet parameterized models after training to different window lengths $w$~days, computed as the sum up to normalized wavenumber $k = 20$ of the time-averaged kinetic energy spectrum, and in units of \unit{\m^4\per\s\squared}. The value for the high resolution $2049 \times 2049$ reference is shown in red, and for the filtered and coarse grained data in dotted red.}\label{fig:resnet_large_scale_ke}
\end{figure}

The large-scale kinetic energy is diagnosed by computing the sum of the energies in the largest wavenumbers, up to and including $k = 20$ (where $k = 1$ corresponds to a wavelength of $4 L$), and the results for all stable ResNet parameterized models are shown in Figure~\ref{fig:resnet_large_scale_ke}. There is a general pattern of decreasing large scale kinetic energy with increasing window length.

\begin{figure}
  \begin{center}
  \includegraphics[height=0.35\textwidth]{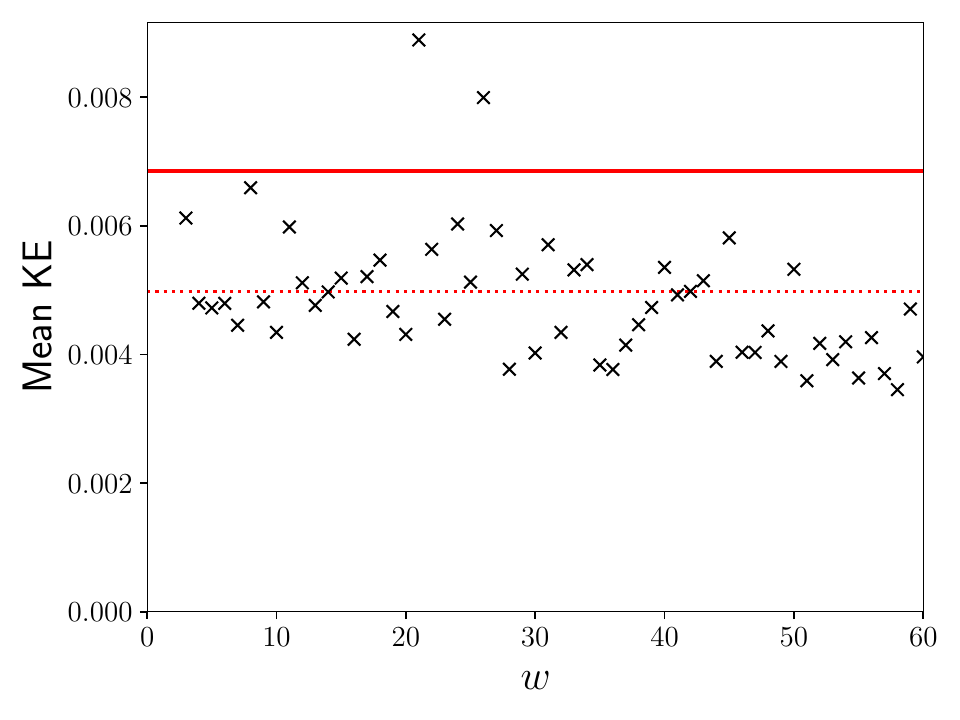}
  \includegraphics[height=0.35\textwidth]{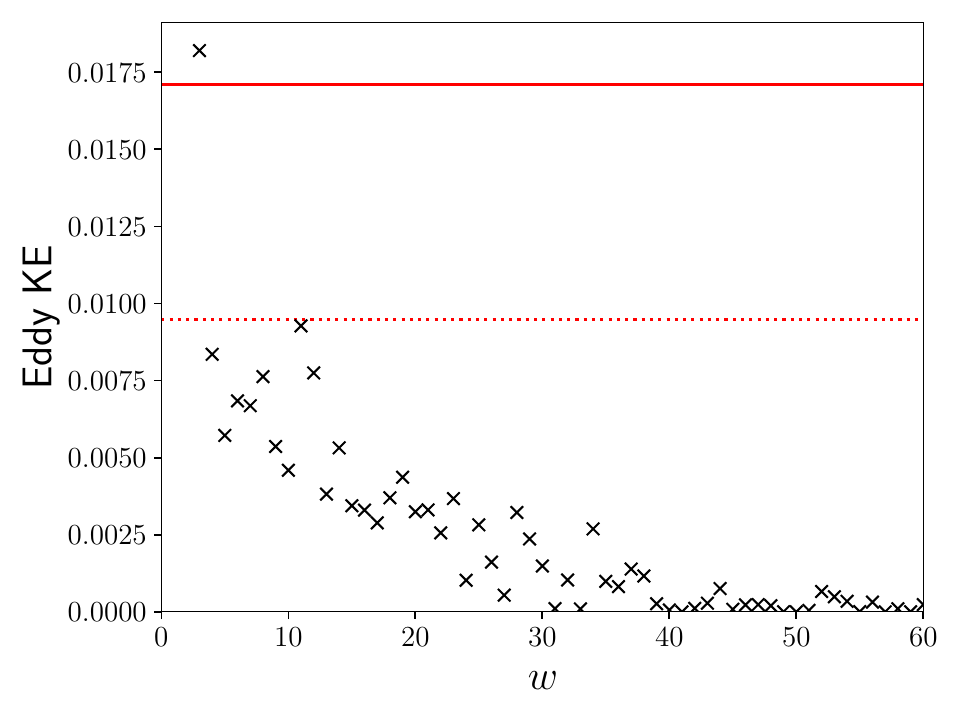}
  \end{center}
  \caption{Domain averaged mean kinetic energy (left) and eddy kinetic energy (right), in units of \unit{\m\squared\per\s\squared}, for the ResNet parameterized models (black crosses) after training to different window lengths $w$~days. The energy for the high resolution $2049 \times 2049$ reference is shown with the red line, and the energy associated with the filtered and coarse grained data is shown with the red dotted line.}\label{fig:resnet_ke}
\end{figure}

Figure~\ref{fig:resnet_ke} shows the domain averaged mean and eddy kinetic energies for each of the ResNet parameterized models, with the mean kinetic energy defined via $K_\text{mean} = \overline{u} \cdot \overline{u} / 2$ and eddy kinetic energy via $K = \overline{u' \cdot u'} / 2$. The value from the high resolution $2049 \times 2049$ reference is also shown. In addition, the energy associated with the Gaussian filtered and coarse grained data is shown via the red dotted line. The mean energies for the ResNet parameterized models generally vary around the mean energy associated with the filtered and coarse grained data, albeit with significant variability, and with a decreasing trend with window length. For the shortest window length the ResNet parameterized model has an eddy kinetic energy which is significantly higher than might be expected given the training set. The eddy kinetic energy then decreases as the window length is increased, until at long window lengths the eddy energy is very low.

The pattern in eddy kinetic energy suggests that there might be a trade-off associated with the training window length. The window length may need to be sufficiently large that the neural network is trained to ensure numerical stability -- numerical instability may need to have sufficient time to develop and influence the loss in order for the neural network to be trained to avoid it. However the window length may also need to be sufficiently short, so that it is within the range of predictability of the reference system.

\begin{figure}
  \begin{center}\begin{tabular}{ccc}
    \includegraphics[height=0.3\textwidth]{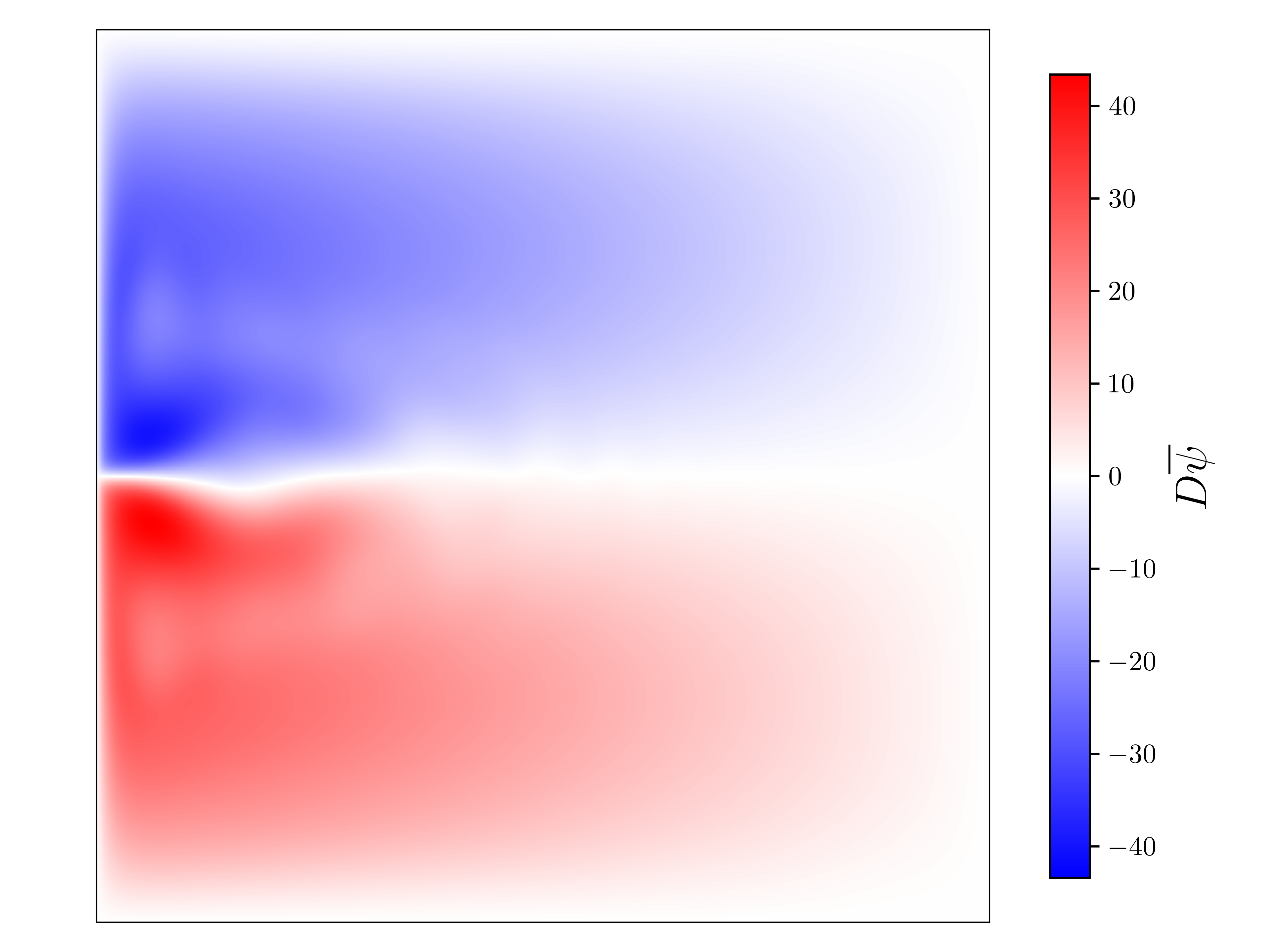}
    & &
    \includegraphics[height=0.3\textwidth]{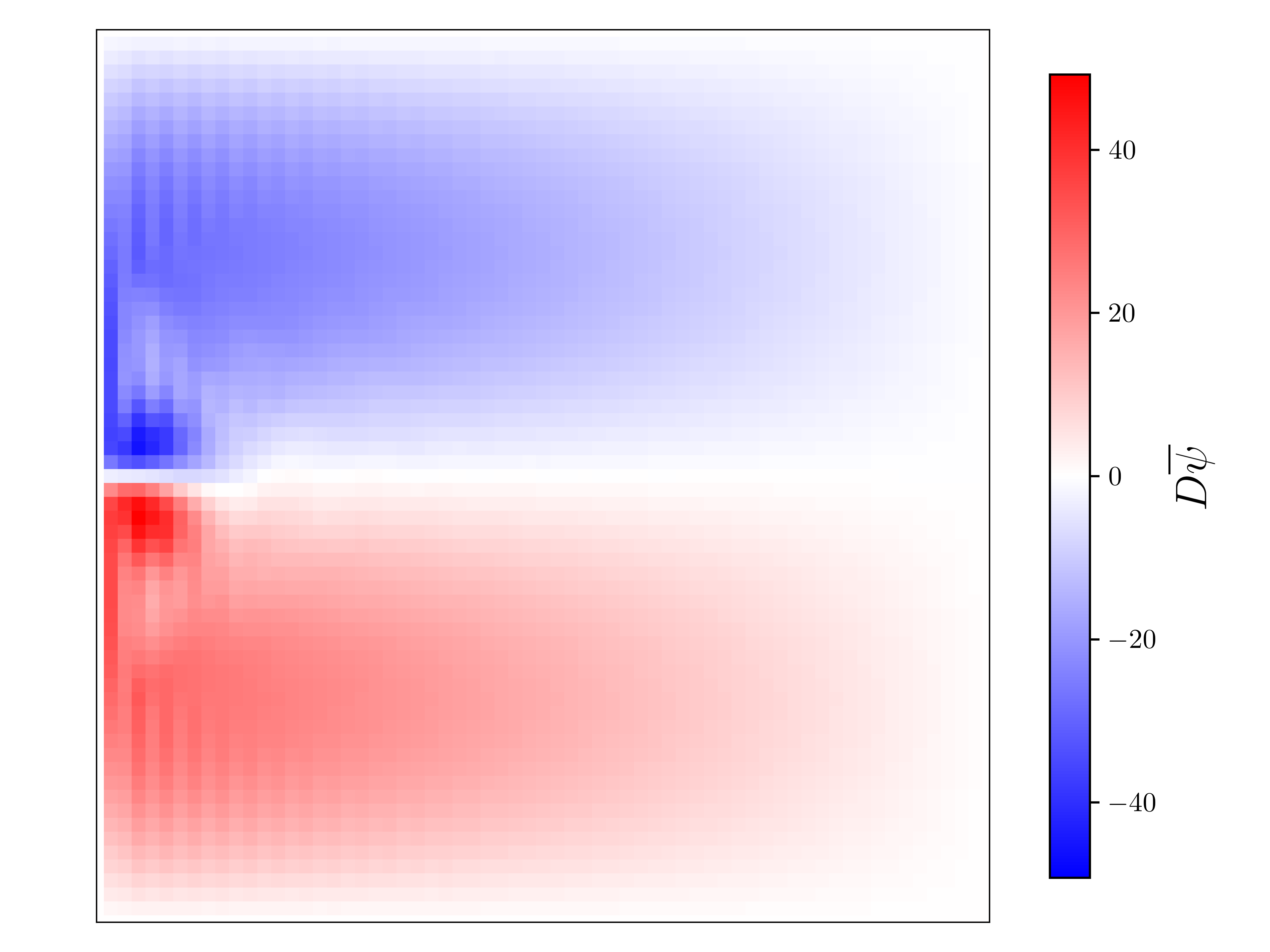} \\
    $2049 \times 2049$ \qquad\qquad
    & &
    $65 \times 65$ \qquad\qquad \\
    & & \\
    \includegraphics[height=0.3\textwidth]{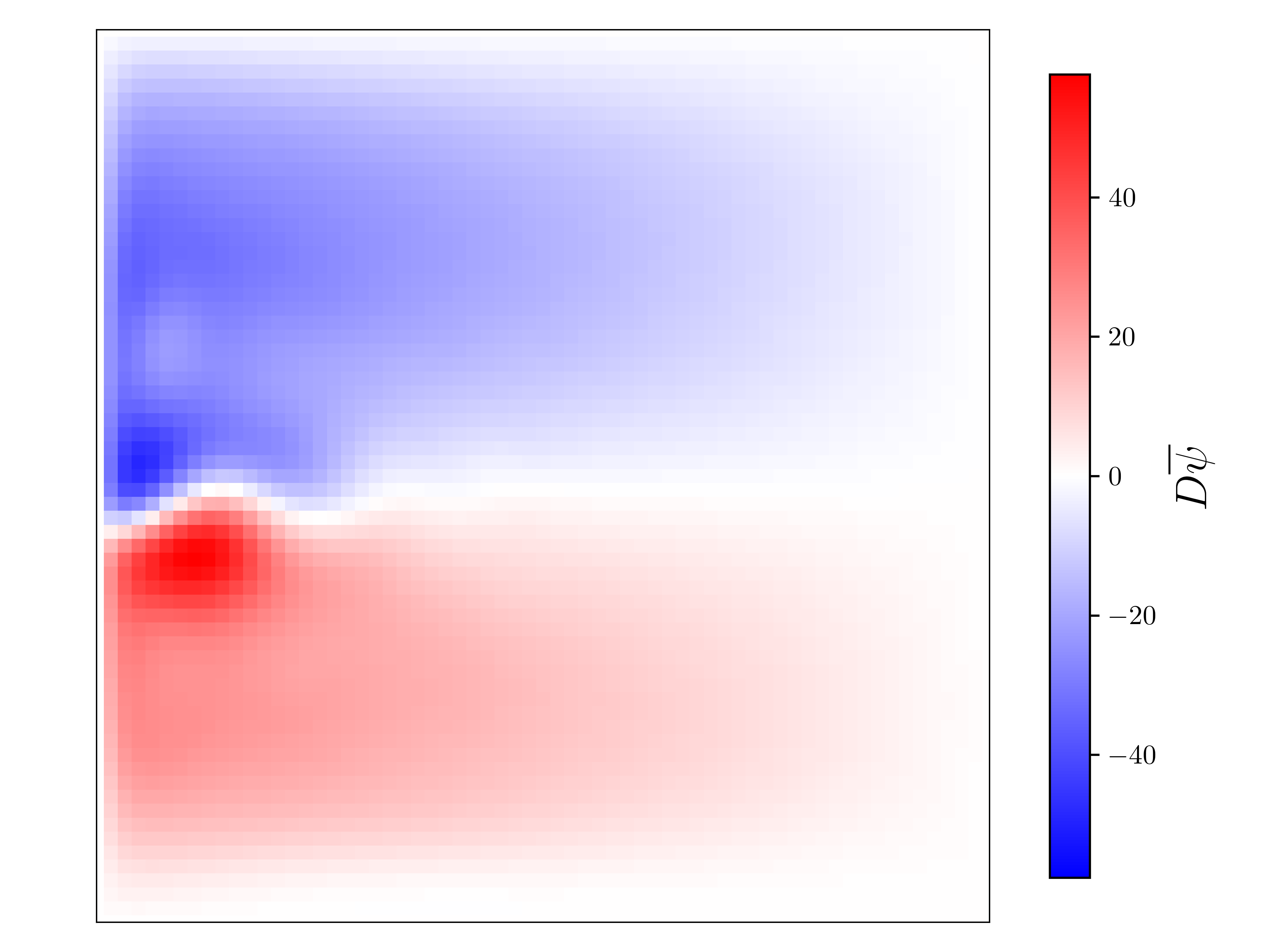}
    & &
    \includegraphics[height=0.3\textwidth]{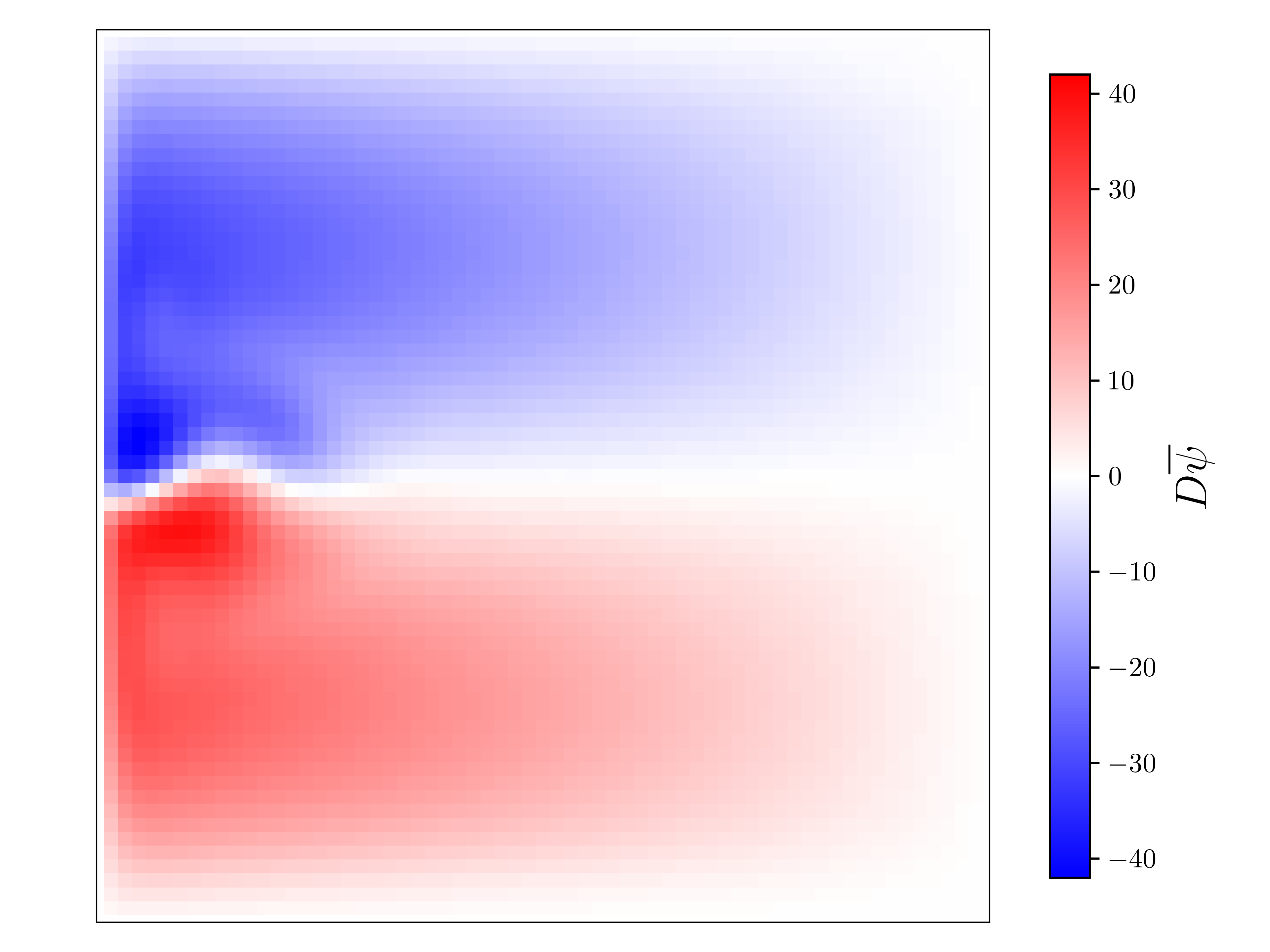} \\
    $65 \times 65$ + ResNet, $w = 3$~days \qquad\qquad
    & &
    $65 \times 65$ + ResNet, $w = 4$~days \qquad\qquad \\
    & & \\
    \includegraphics[height=0.3\textwidth]{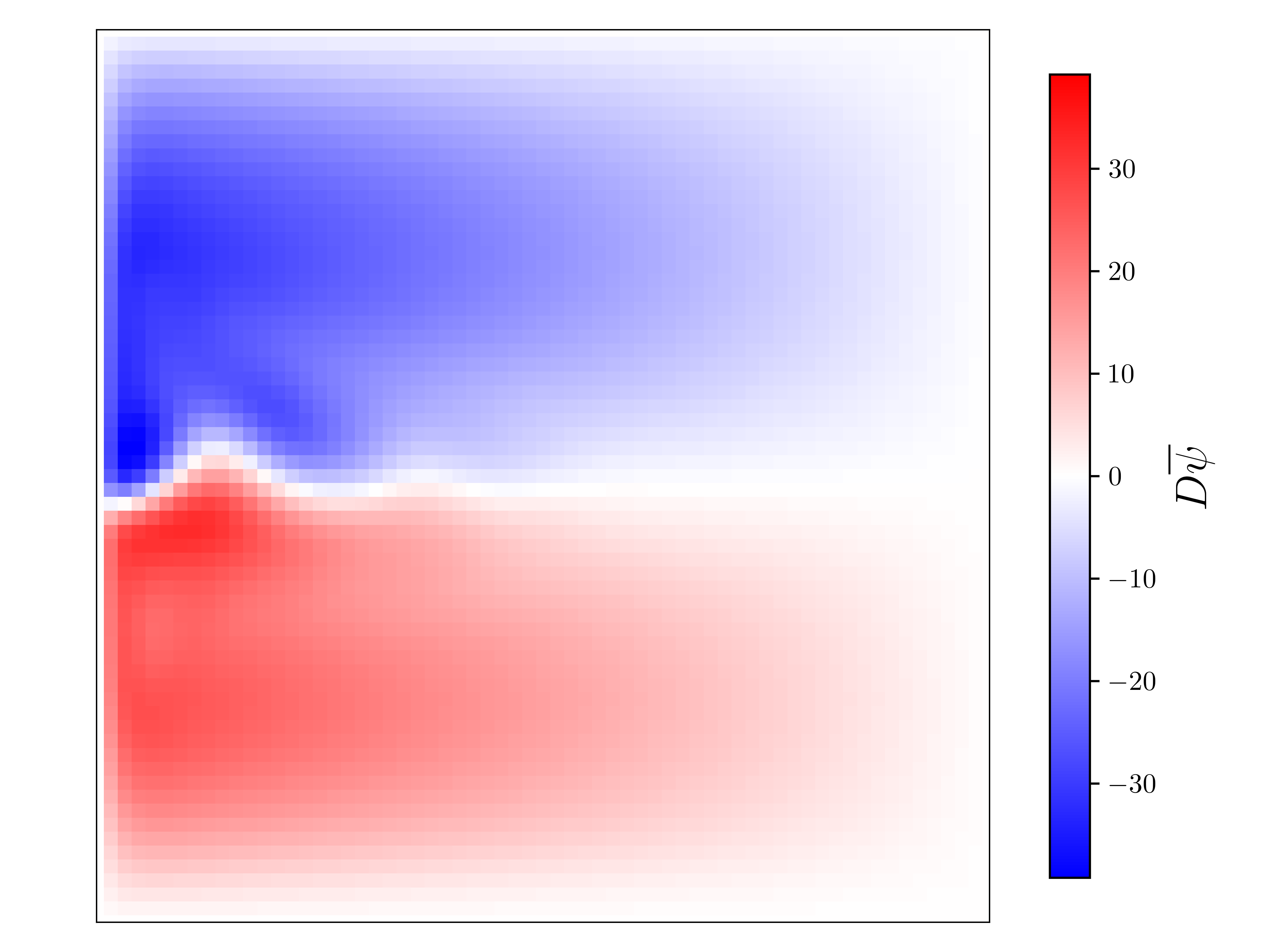}
    & &
    \includegraphics[height=0.3\textwidth]{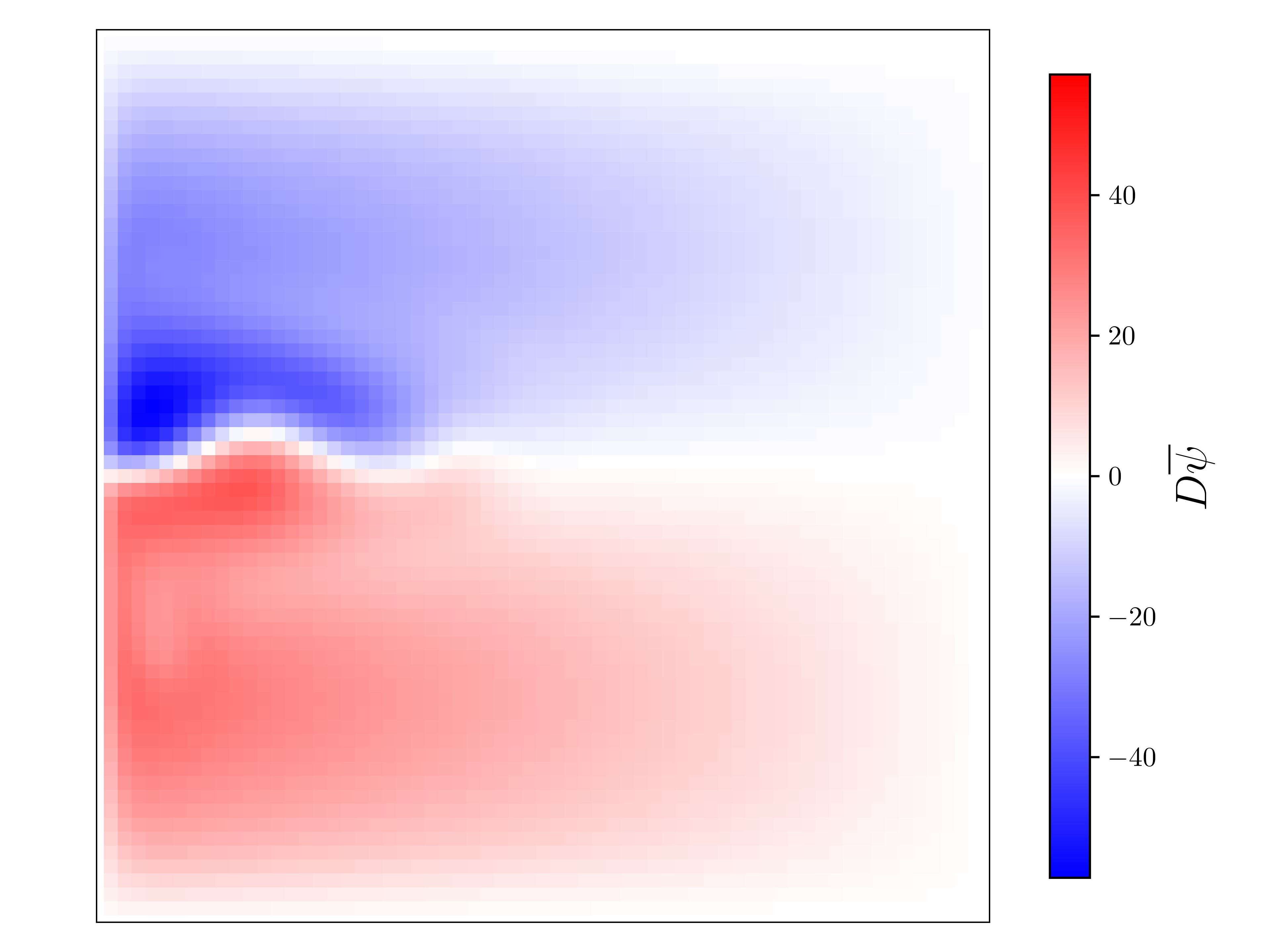} \\
    $65 \times 65$ + ResNet, $w = 7$~days \qquad\qquad
    & &
    $65 \times 65$ + ResNet, $w = 8$~days \qquad\qquad \\
    \includegraphics[height=0.3\textwidth]{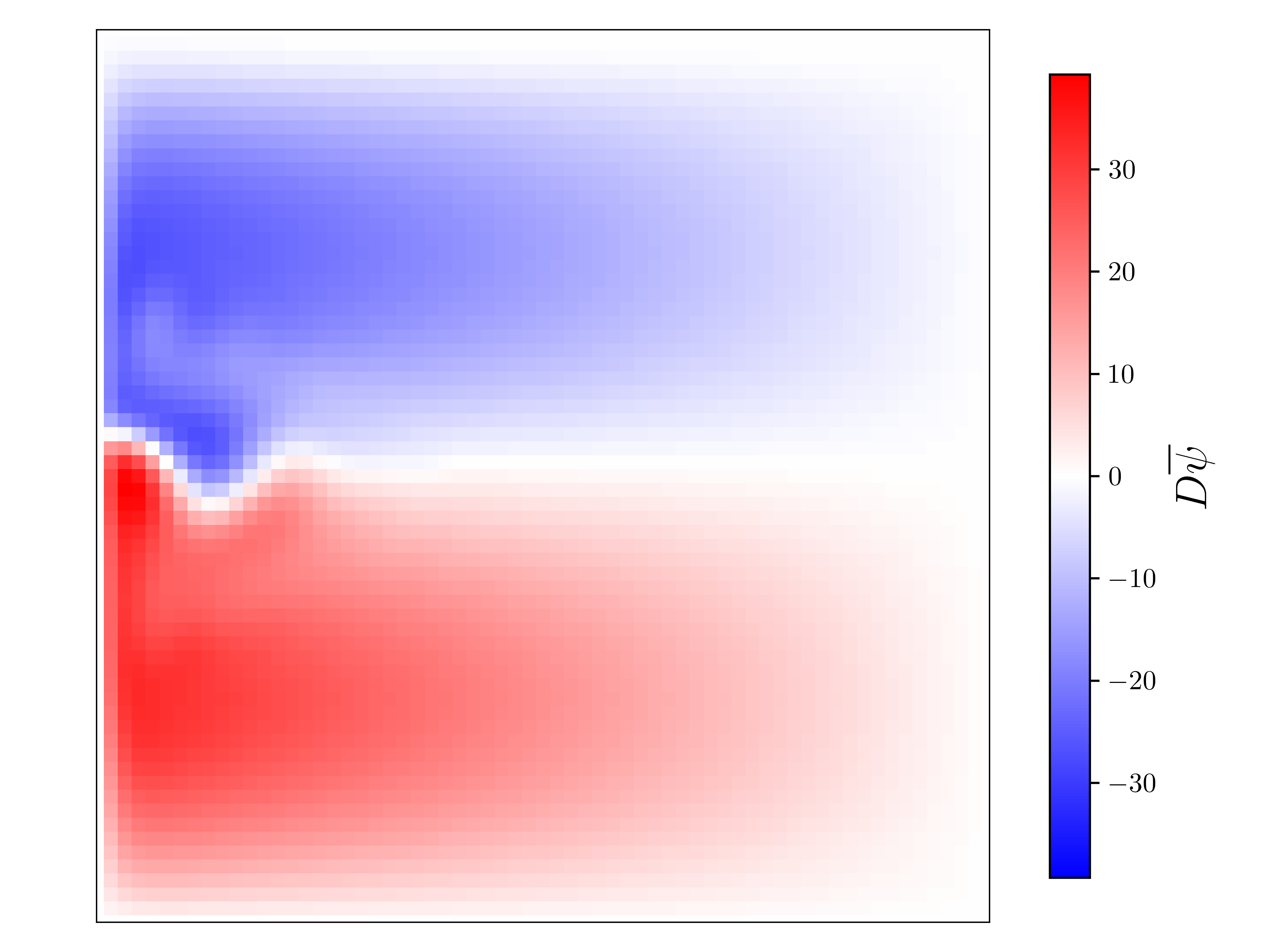}
    & &
    \includegraphics[height=0.3\textwidth]{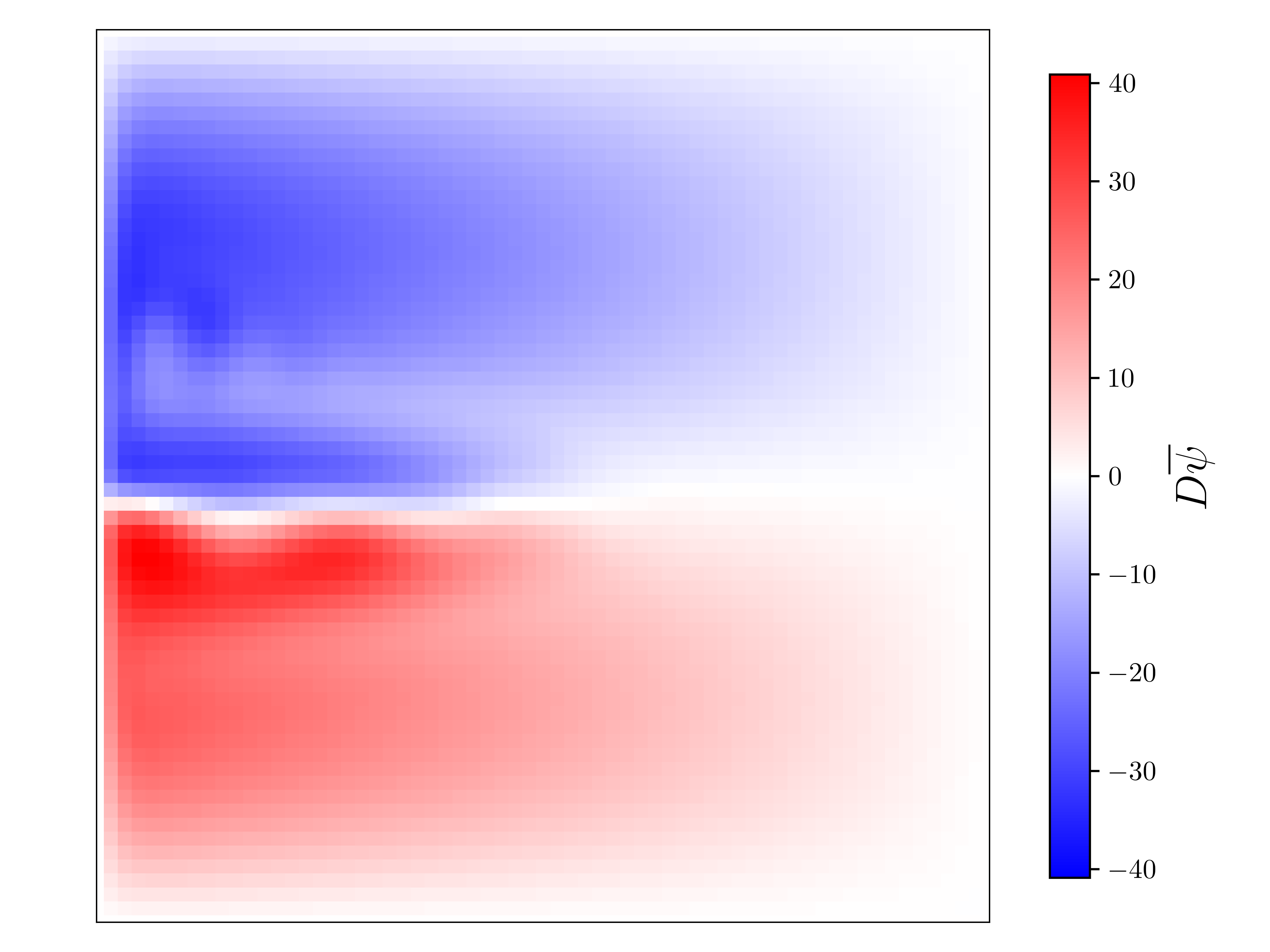} \\
    $65 \times 65$ + ResNet, $w = 16$~days \qquad\qquad
    & &
    $65 \times 65$ + ResNet, $w = 17$~days \qquad\qquad
  \end{tabular}\end{center}
  \caption{Mean transport stream function for the high resolution $2049 \times 2049$ reference, the coarse resolution $65 \times 65$ control, and for the ResNet parameterized models after training up to different window lengths, in units of Sv. Note that different color scales are used.}\label{fig:resnet_mean_psi}
\end{figure}

Figure~\ref{fig:resnet_mean_psi} shows the mean stream function for the ResNet parameterized model with a number of different window lengths, with the high resolution $2049 \times 2049$ reference and coarse resolution $65 \times 65$ control for comparison. The eastward extent of the mean jet is generally improved compared to the coarse resolution $65 \times 65$ control, although there is significant variability, and at longer window lengths there are cases with more limited extent (not shown).

\begin{figure}
  \begin{center}\begin{tabular}{ccc}
    \includegraphics[height=0.3\textwidth]{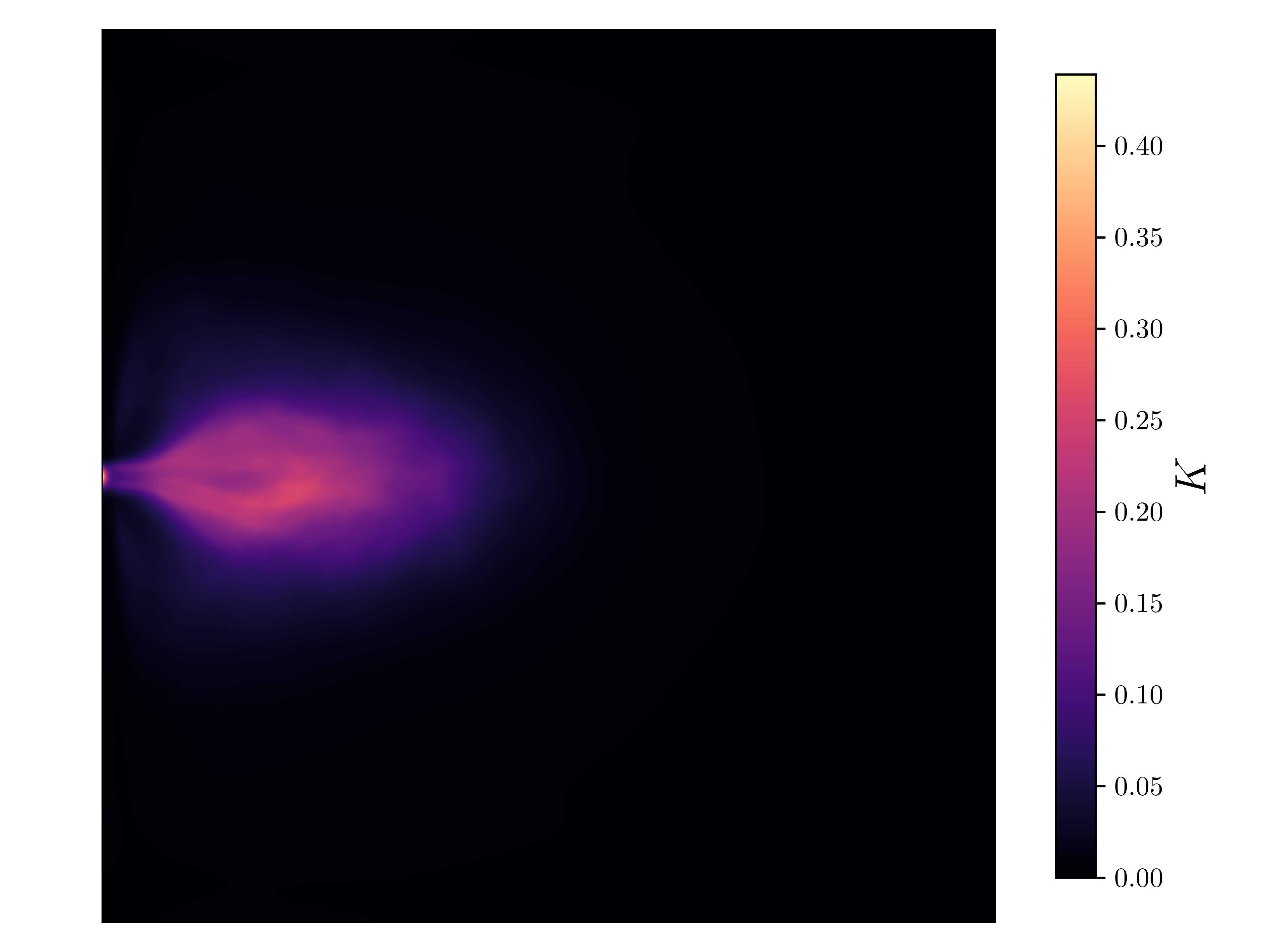}
    & &
    \includegraphics[height=0.3\textwidth]{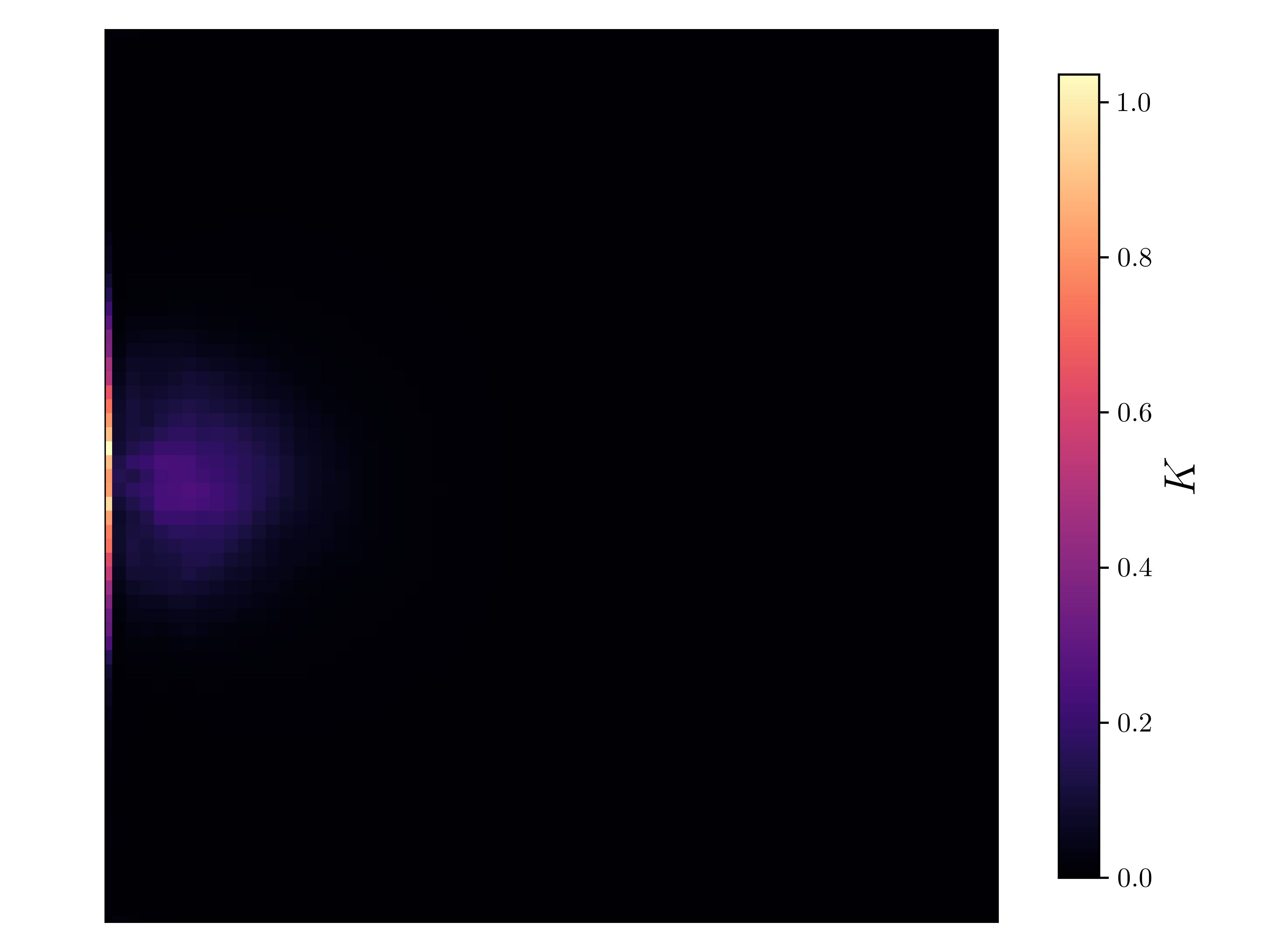} \\
    $2049 \times 2049$ \qquad\qquad
    & &
    $65 \times 65$ \qquad\qquad \\
    & & \\
    \includegraphics[height=0.3\textwidth]{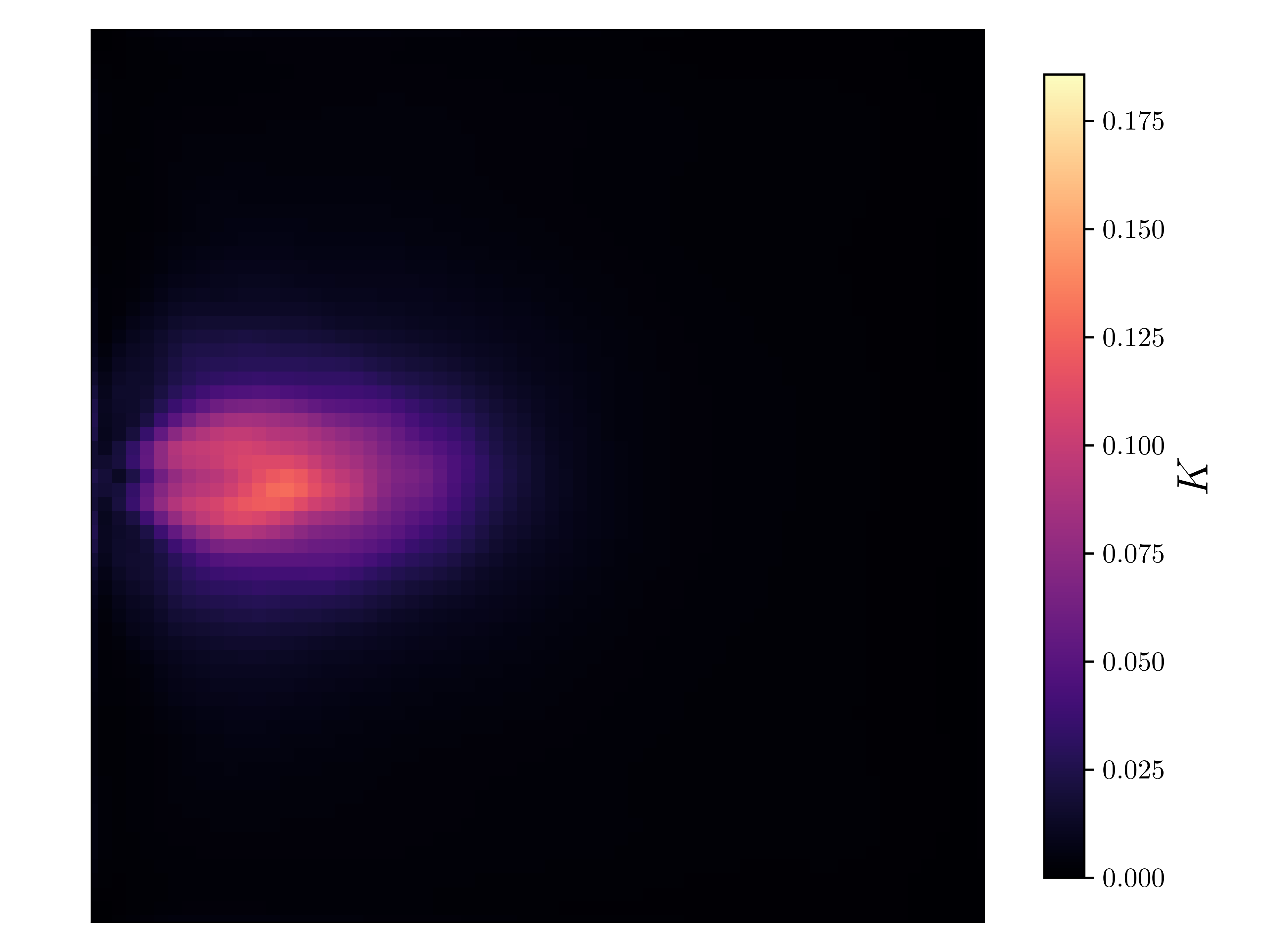}
    & &
    \includegraphics[height=0.3\textwidth]{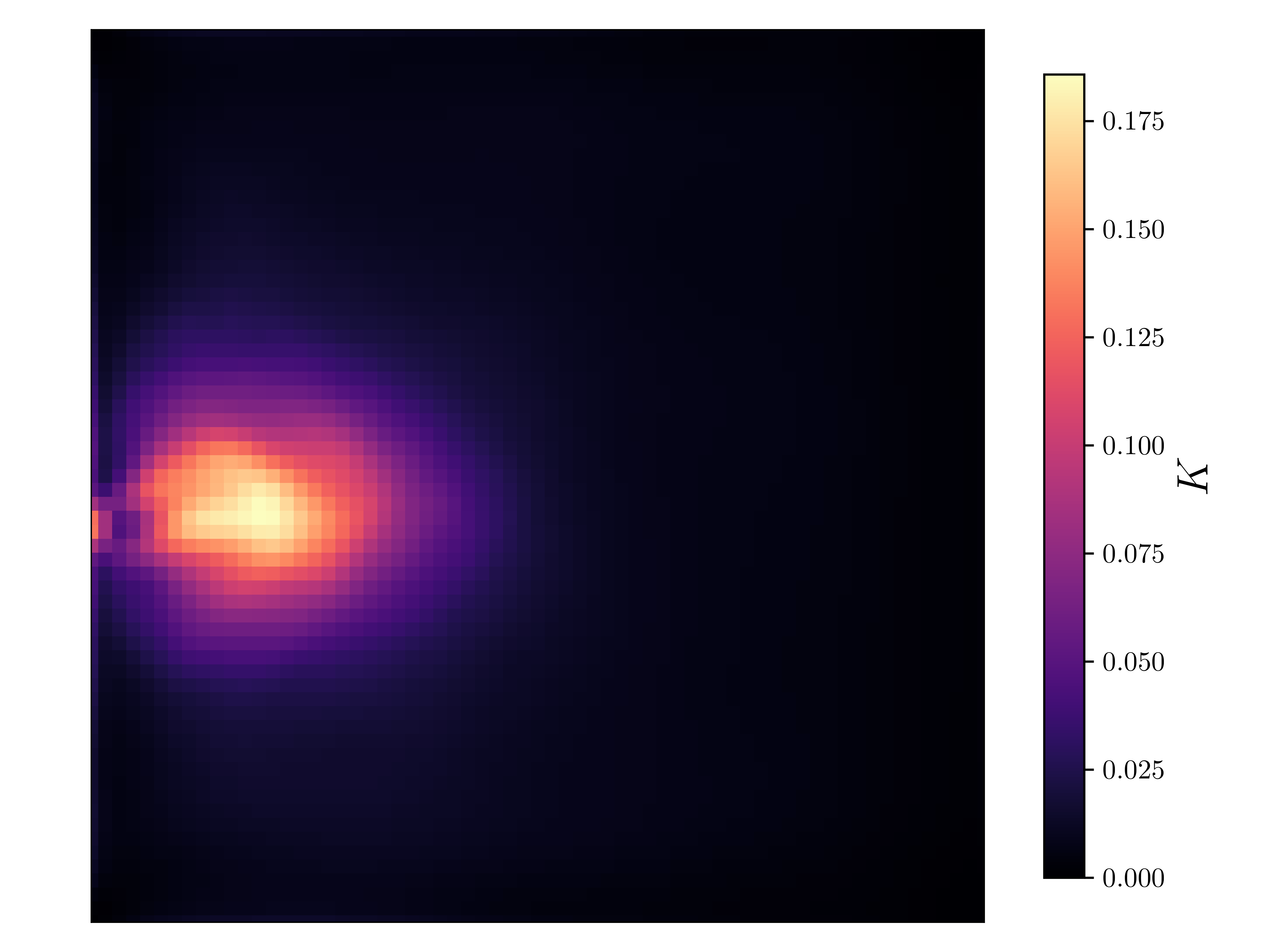} \\
    $2049 \times 2049$, filtered + coarse grained \qquad\qquad
    & &
    $65 \times 65$ + ResNet, $w = 3$~days \qquad\qquad \\
    & & \\
    \includegraphics[height=0.3\textwidth]{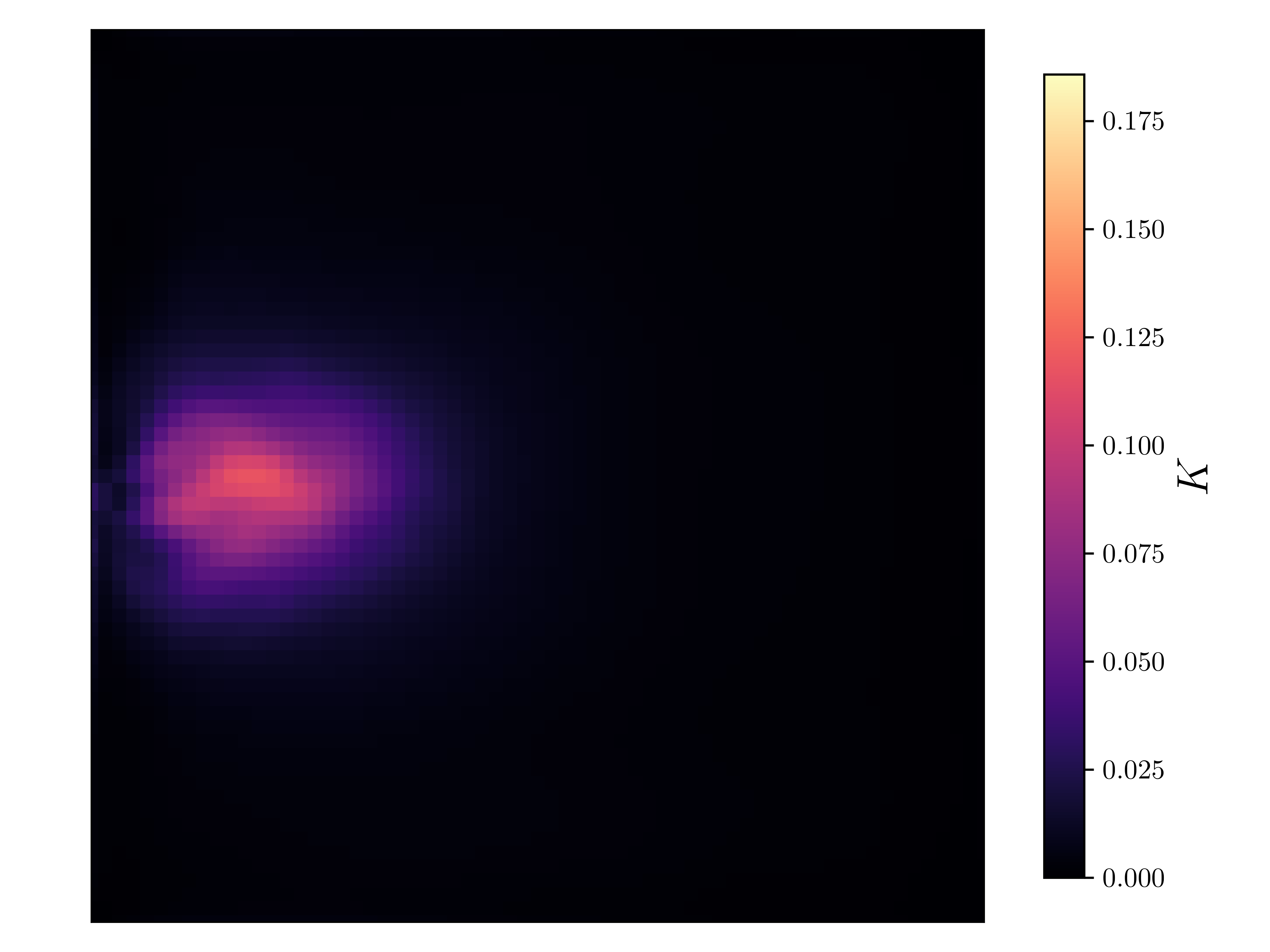}
    & &
    \includegraphics[height=0.3\textwidth]{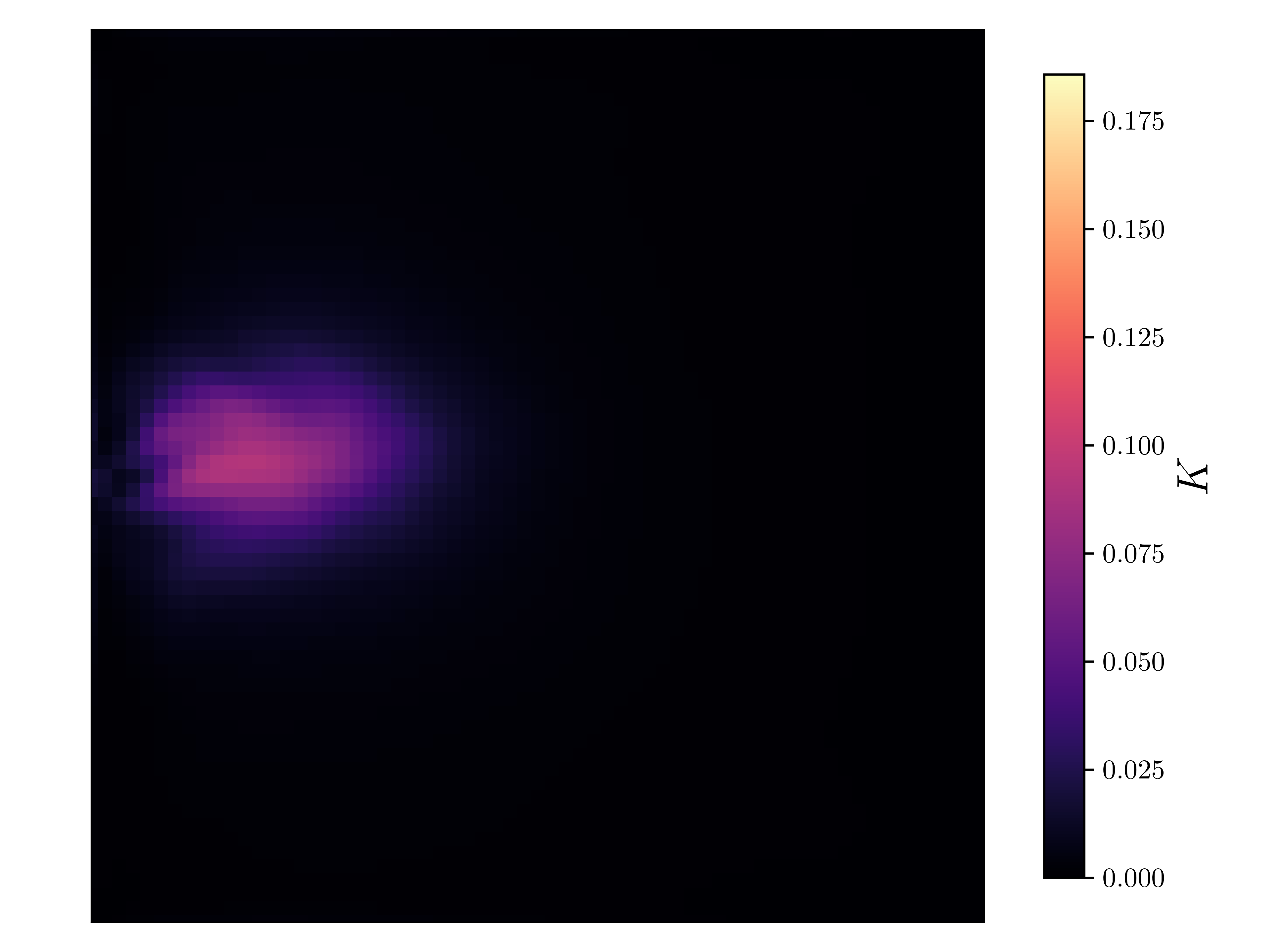} \\
    $65 \times 65$ + ResNet, $w = 4$~days \qquad\qquad
    & &
    $65 \times 65$ + ResNet, $w = 5$~days \qquad\qquad
  \end{tabular}\end{center}
  \caption{Eddy kinetic energy for the high resolution $2049 \times 2049$ reference, the coarse resolution $65 \times 65$ control, associated with the filtered and coarse grained data, and for the ResNet parameterized model after training up to different window lengths, with units \unit{\m\squared\per\s\squared}. Different color scales are used for the reference and control (upper two panels), and matching color scales are used for the remaining panels.}\label{fig:resnet_eke}
\end{figure}

Figure~\ref{fig:resnet_eke} shows the eddy kinetic energy for the ResNet parameterized model trained up to window lengths of $w =3$, $w = 4$, and $w = 5$~days, with the high resolution $2049 \times 2049$ reference for comparison. The eddy kinetic energy associated with the filtered and grained data is also shown. These ResNet parameterized models exhibits significant internal variability, with a reasonable match with the filtered coarse grained data, both in terms of location and magnitude.

The general pattern in the ResNet parameterized system is that, once the neural network is trained up to a window size sufficient for numerical stability, but for window lengths that are not too long, the system has an improved mean flow state in terms of the mean jet extension, and improved internal variability The improvement over the non-parameterized coarse resolution reference is dramatic, with the parameterized system being significantly smoother and with modest near grid scale noise, but also having resolved kinetic energy which is, for shorter window lengths, comparable with the energy in the filtered and coarse grained data. That is, online learning at shorter window lengths leads to a parameterized system which looks similar to a filtered and coarse grained view of the original high resolution system used to generate the training data.

\section{Generalizability tests}\label{sect:generalizability}

In the preceding section the trained ResNets have been assessed only against their ability to simulate dynamics within ranges of parameters that appear in the dataset. This leads to concerns regarding generalizability. For example, one can ask whether the neural networks can be applied to more general configurations, and to explore questions which could not have been explored using the training set alone.

Note that, to an extent, the discussion in the preceding section has already considered an `out-of-sample' problem \emph{in time}. The neural networks were trained using simulations only over limited windows, but then prognostic calculations over much longer time periods were considered. In this section two further more directly out-of-sample cases are considered: one exploring a modification to external parameters, and one exploring symmetry preservation, each applied for the ResNet architecture.

\subsection{Modified external parameters}

For an out-of-sample test, the wind stress is rotated by $\theta = \pi / 6$, meaning that \eqref{eqn:wind_stress} is modified to
\begin{equation*}
  \tau = \tau_0 \cos \left( \frac{ \pi ( y \cos \theta - x \sin \theta )}{L} \right) \left( \begin{array}{c} \cos \theta \\ \sin \theta \end{array} \right).
\end{equation*}
The wind stress magnitude is also increased to $\tau_0 = 0.15$~\unit{\N\per\m\squared}. All other parameters are left unchanged, and the trained ResNets are used unmodified. ResNets trained up to window lengths $w = 1$ to $w = 24$~days are considered. There is a choice to be made as to whether the non-dimensional scaling parameter $\alpha_\text{output}$ should be increased for this case (since it is defined in terms of $\tau_0$, and $\tau_0$ has been increased). Here we \emph{do} increase $\alpha_\text{output}$ to reflect the increased maximum wind stress magnitude.

In this case the three shortest window lengths at $w = 1$, $w = 2$, and $w = 3$~days led to numerical instability. All other cases led to stable integrations for both the $12$~year spinup, and during a further simulated $12$~years during which time-averaged diagnostics were computed.

\begin{figure}
  \begin{center}\begin{tabular}{ccc}
    \includegraphics[height=0.3\textwidth]{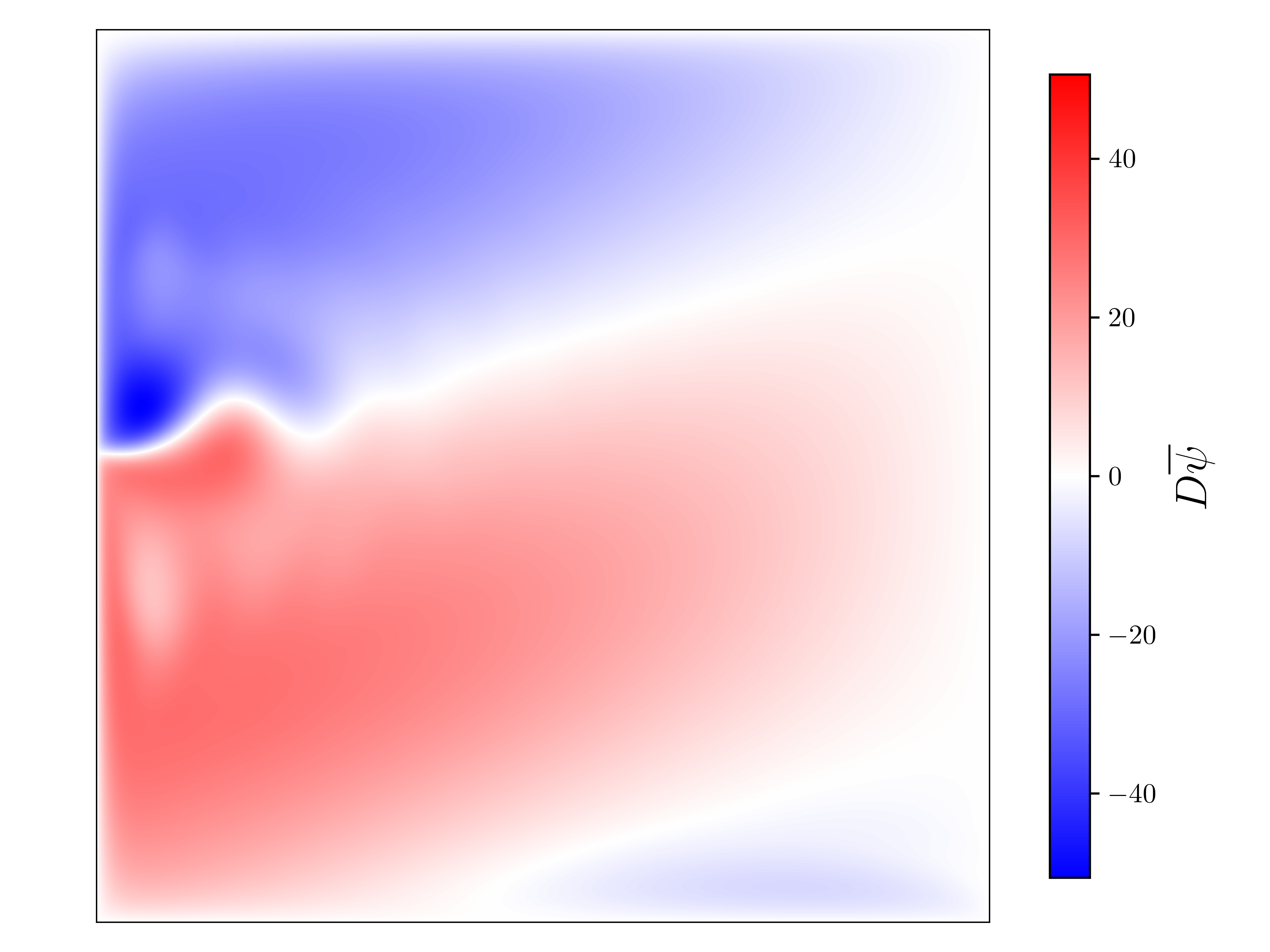}
    & &
    \includegraphics[height=0.3\textwidth]{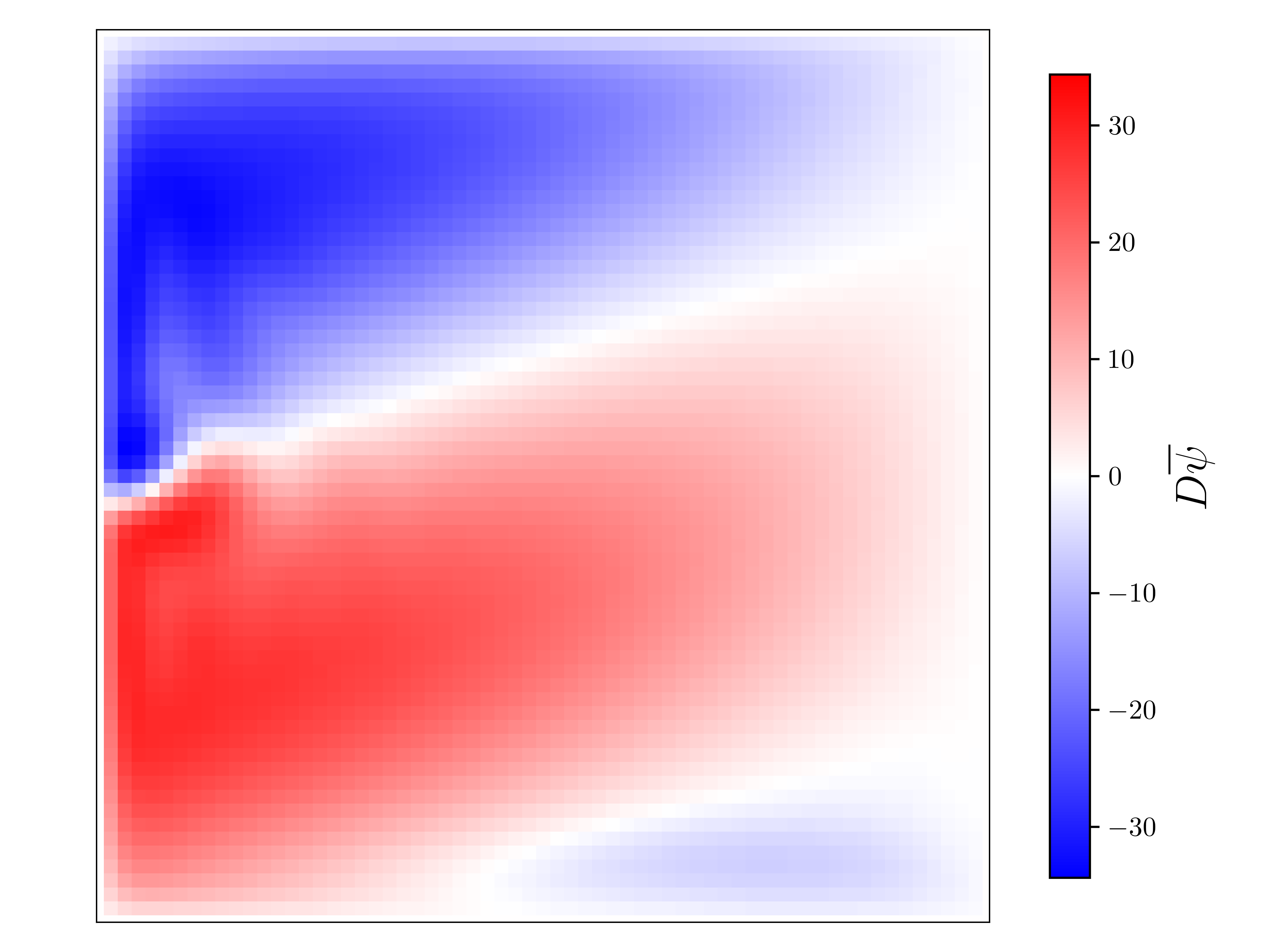} \\
    & & \\
    $2049 \times 2049$ \qquad\qquad
    & &
    $65 \times 65$ + ResNet, $w = 4$~days \qquad\qquad
  \end{tabular}\end{center}
  \caption{Mean transport stream function, in units of Sv, for the out-of-sample test with a rotated wind stress and increased maximum wind stress magnitude. Left: The high resolution $2049 \times 2049$ reference. Right: The ResNet parameterized model, after training up to $w = 4$~days. Note that different color scales are used.}\label{fig:coarse_mean_psi_oos}
\end{figure}

Figure~\ref{fig:coarse_mean_psi_oos} shows the mean transport stream function for the reference calculation, and for the ResNet parameterized model after training up to a window length of $w = 4$~days. It is clear that features of the high resolution $2049 \times 2049$ reference flow are qualitatively reproduced, including an appropriate tilt of the mean jet, and an additional circulation in the southeast corner of the domain, although there is some difference in magnitude.

\begin{figure}
  \begin{center}\begin{tabular}{c}
    \includegraphics[height=0.3\textwidth]{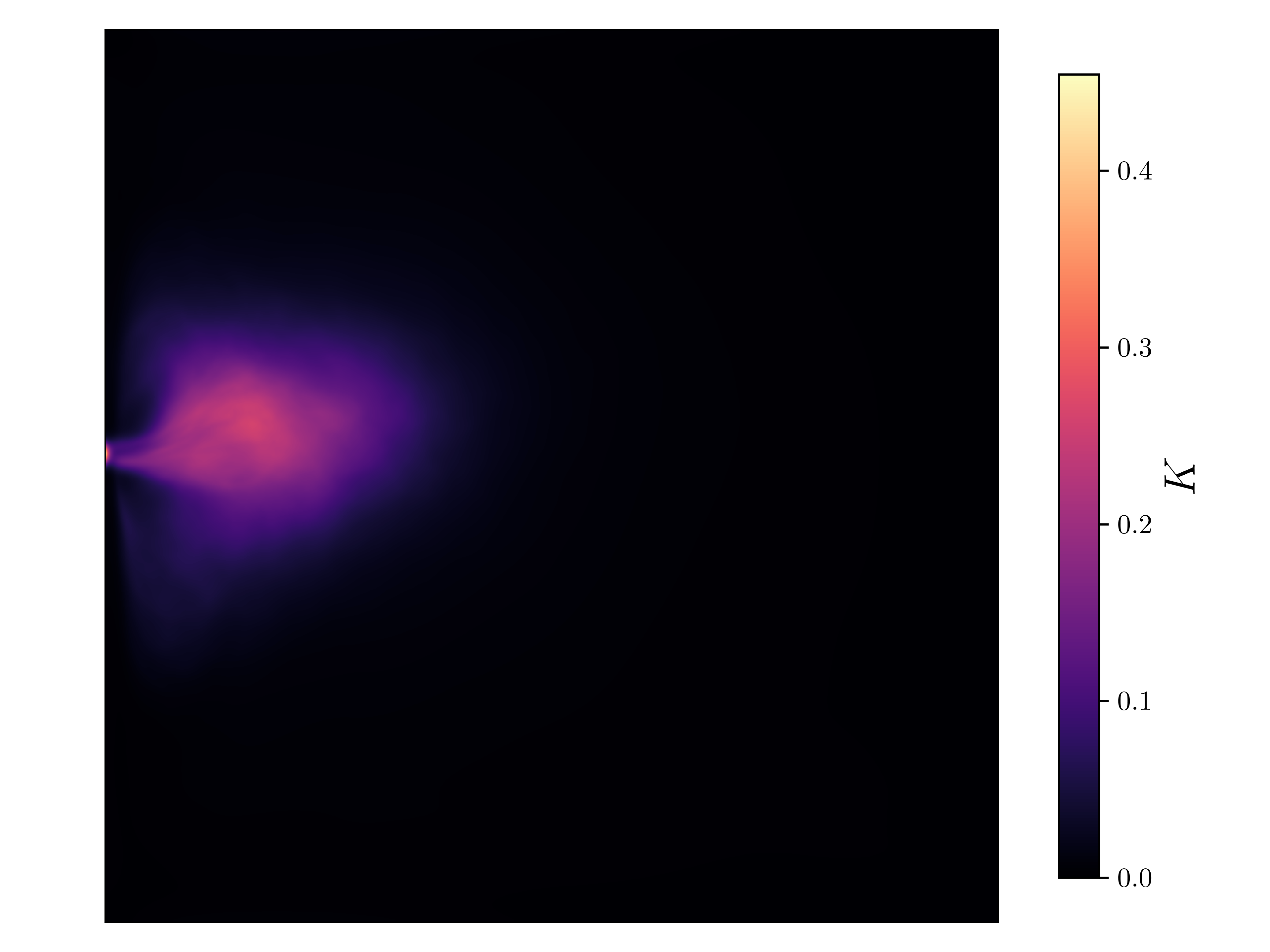} \\
    $2049 \times 2049$ \qquad\qquad
    \\
  \end{tabular}
  \begin{tabular}{ccc}
    \includegraphics[height=0.3\textwidth]{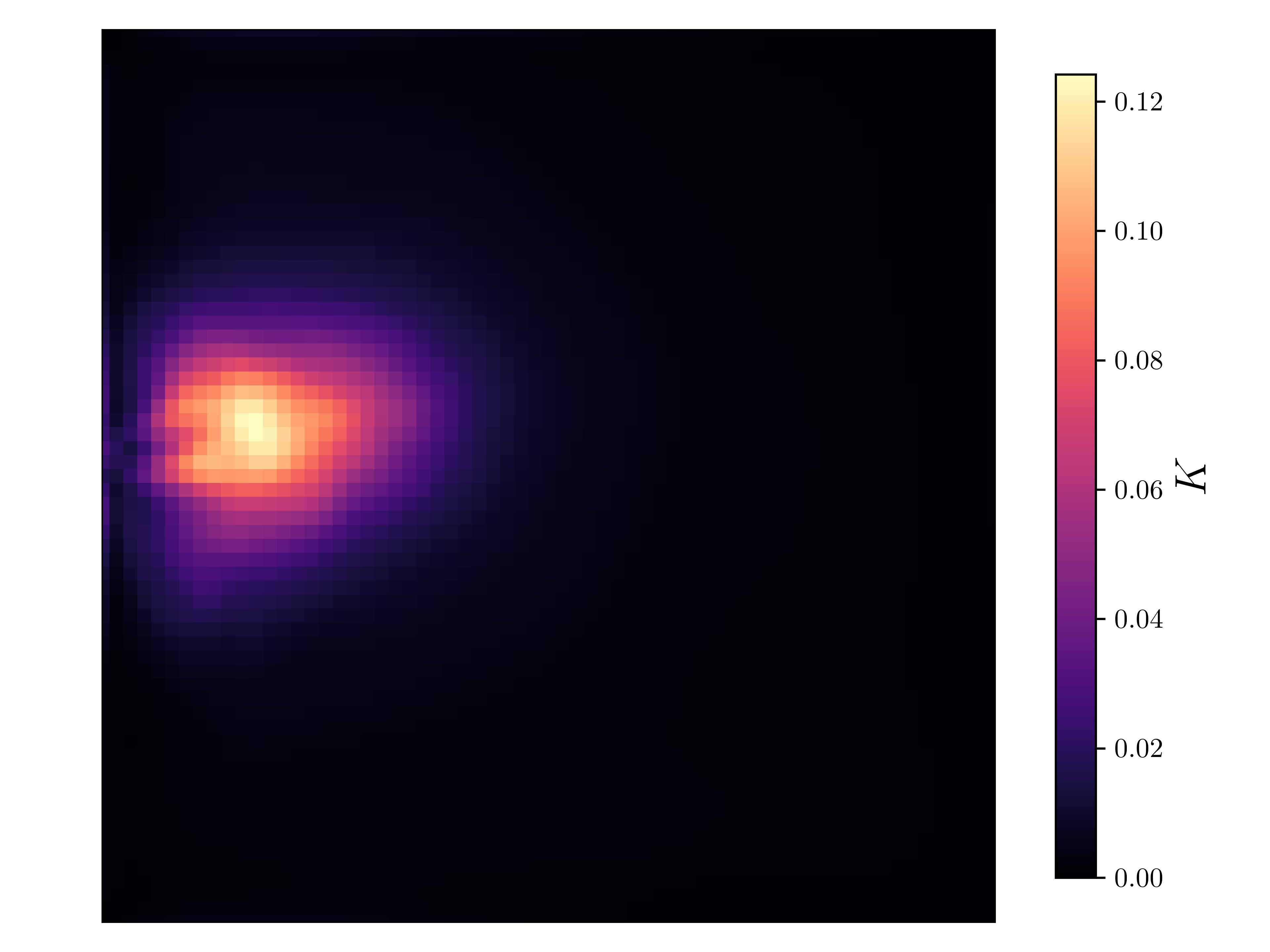}
    & &
    \includegraphics[height=0.3\textwidth]{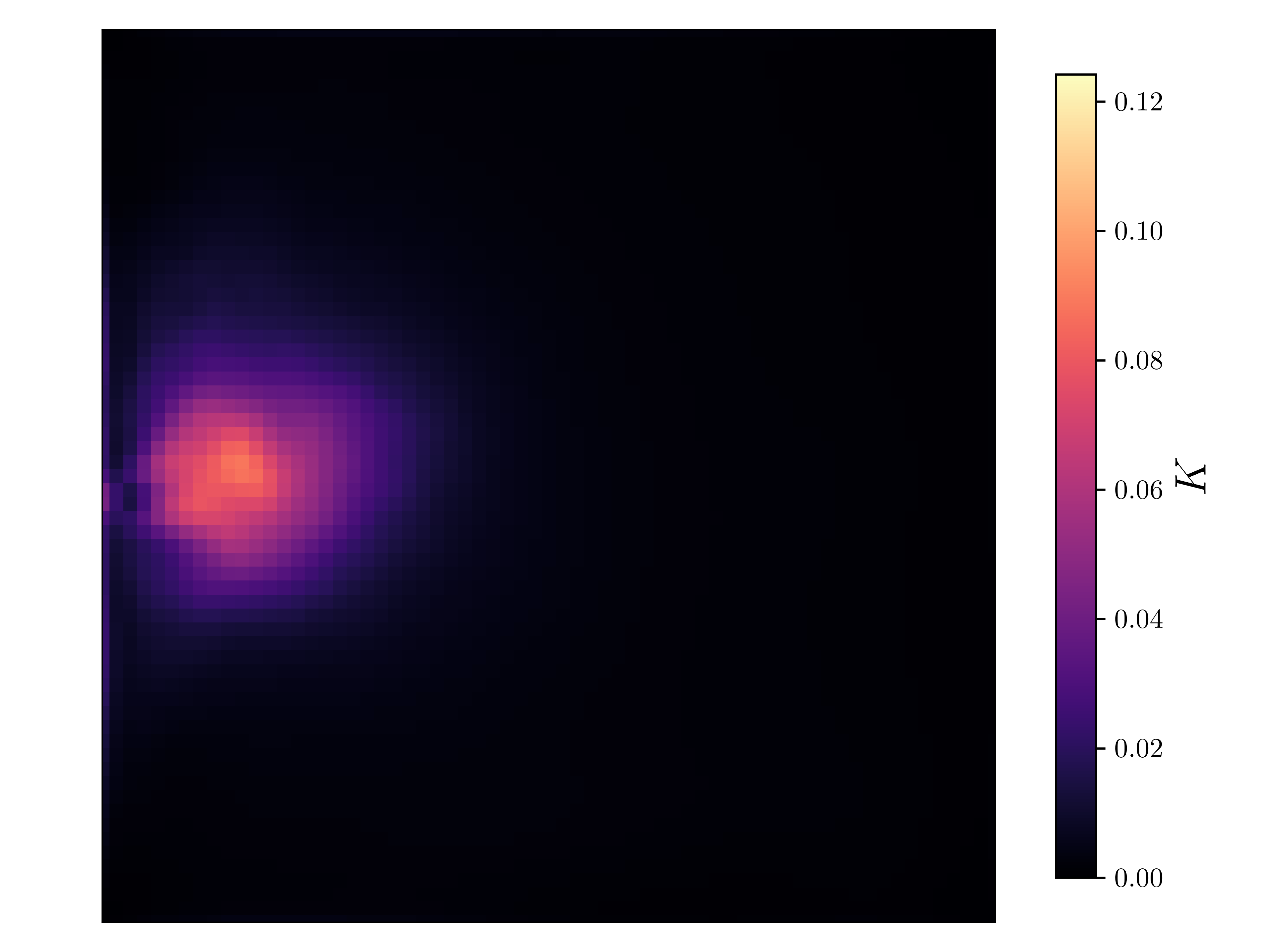} \\
    $2049 \times 2049$, filtered and coarse grained \qquad\qquad
    & &
    $65 \times 65$ + ResNet, $w = 4$~days \qquad\qquad
  \end{tabular}\end{center}
  \caption{Eddy kinetic energy in the out-of-sample test with a rotated wind stress and increased maximum wind stress magnitude, for the high resolution $2049 \times 2049$ reference, associated with the filtered and coarse grained data, and for the ResNet parameterized model trained up to $w = 4$~days, with units \unit{\m\squared\per\s\squared}. Note that a different color scale is used for the reference, and that matching color scales are used for the remaining two panels.}\label{fig:resnet_eke_4_oos}
\end{figure}

Figure~\ref{fig:resnet_eke_4_oos} shows the eddy kinetic energy for the ResNet parameterized model trained up to a window length of $w = 4$~days, with the high resolution $2049 \times 2049$ reference and the eddy kinetic energy as diagnosed from the filtered and coarse grained data for comparison. The ResNet parameterized model again has a notable degree of variability in the region of the separating jet, although is somewhat less energetic than the filtered and coarse grained data. At $w = 17$, $w = 18$, $w = 19$, and $w = 23$~days~days a spurious large scale eddy energy is seen, perhaps suggesting some degree of instability (not shown).

\subsection{Symmetry}

Convolutional neural networks are a natural architecture for embedding translational symmetry. However the barotropic vorticity equation exhibits other symmetries, and in particular the system is symmetric under a coordinate inversion, $x \rightarrow -x$, together with a change of sign for $\zeta$, $\psi$, $\beta$, and the wind stress curl term. For the neural network parameterized models as applied here, the neural network maps a vorticity to a vorticity tendency. In order for the neural network to exhibit the appropriate symmetry it would need to be invariant under a combination of a horizontal flip and sign change to both its input and output -- a non-trivial symmetry which the neural networks do not intrinsically respect.

The invariance of the ResNet to a reflection of the $x$-coordinate was tested, changing the sign of $\beta$ and the sign of the wind stress curl term, and otherwise using the ResNets obtained after training up to $w = 24$~days unmodified. The performance of the ResNet parameterized models in this case is poor. Only window lengths $w = 7$, $w = 9$, $w = 14$, and $w = 17$~days led to stable integrations for a full $24$ year calculation. Unphysical features are apparent, such as significant deflection of the mean jet (for $w = 7$, $w = 9$, and $w = 14$~days), spurious mean circulations in the eastern corners (for $w = 7$, $w = 9$, and $w = 14$~days), and significant eddy energy on the eastern boundary (particularly for $w = 9$, $w = 14$, and $w = 17$~days) (not shown).

It is clear that the ResNet parameterized models do not generalize after applying a symmetry transformation under which the reference model, used to generate the training data, is invariant.

\section{Summary and conclusions}\label{sect:conclusions}

This article has considered the application of online learning to a highly idealized barotropic ocean gyre model, seeking to assess the performance of neural network parameterized models at coarse resolution. It is found that, while stability and performance is variable, when suitably configured and trained, such models can lead to reasonable mean flows and intrinsic variability, and can lead to stable solutions despite the very low explicit viscosity used. With the residual neural network there appears to be a trade-off when choosing the training window length between stability, obtained at longer window lengths, versus increased variability, obtained at shorter window lengths.

Out-of-sample tests have been briefly considered. A test with a tilted wind stress of increased magnitude suggests some degree of generalizability, with a reasonable representation of the mean flow and variability. However a symmetry test led to a clear failure -- the neural network parameterized results changed when applying a symmetry transformation under which the dynamics should be invariant, leading to instability or worse results.

While the ResNet applied here is relatively modest is size, evaluation of the network still has a very significant relative computational cost. With a single CPU process, timestepping with network evaluation is $\sim 100$ times the cost of timestepping alone -- noting that the dynamical model considered here is simple, small, and efficient. Evaluation of convolutional layers is parallelizable, but detailed performance analysis on a GPU is complicated by the very small size of the coarse resolution dynamical model. It nevertheless seems important to investigate whether smaller and cheaper neural networks can also be applied to this problem -- since it is essential that a neural network parameterized model should compete, in terms of performance, with simply increasing dynamical model resolution. Note the recent work of \citeA{srinivasan2024} reporting offline learning results with much smaller neural networks.

It should also be noted that online learning as applied here may learn not only from the supplied training data, but also from the coarse resolution numerics. That is, the system must learn not only to add missing physics, but also to correct any numerical problems added by the use of coarse resolution. While the trained neural networks might be more immediately useful for constructing emulators, it is unclear whether a neural network, trained using online learning when embedded within one model, might be suitable for embedding within any other model.

In principle, with a sufficiently expressive neural network, one might hope to simply use the neural network to model the full dynamics. However this would discard our existing physical knowledge. Online learning bridges the gap between process-based knowledge, which has been studied and developed over the course of decades, with machine learning techniques, allowing machine learning to be applied to those parts of a problem where our knowledge is lacking.

\section*{Open Research Section}
Calculations in this article make use of bt\_ocean, which is available at \linebreak https://github.com/jrmaddison/bt\_ocean. The version of bt\_ocean as used in this article is at \citeA{bt_ocean_a}, and scripts are at \citeA{online_learning_scripts_b}. Trained neural networks and other model data are at \citeA{online_learning_data}. Note that a previous draft of this article additionally made use use of a convolutional neural network which erroneously included the use of an activation function in its final linear layer -- the referenced data include this case.

\acknowledgments

JRM acknowledges helpful comments from Martin Brolly.

For the purpose of open access, the author has applied a Creative Commons Attribution (CC BY) licence to any Author Accepted Manuscript version arising from this submission.

\bibliography{references}

\end{document}